\documentclass[useAMS,twocolumn]{mn2e}
\usepackage{amssymb}
\usepackage{graphicx}
\usepackage{amsmath}
\usepackage{epsfig}

\def\MSUN{\rm M_{\odot}}

\begin{document}
\title[Radiation driven outflow in AGNs]{Radiation driven outflow in active galactic nuclei: the feedback effects of scattered and reprocessed photons}
\author[C. Liu et al. ]{Chao Liu$^{1,2}$, Feng Yuan$^{1}$\thanks{email: fyuan@shao.ac.cn}, Jeremiah P. Ostriker$^{3,4}$, Zhaoming Gan$^{1}$, Xiaohong Yang$^{1,5}$\\
$^1$Key Laboratory for Research in Galaxies and Cosmology, Shanghai Astronomical Observatory, \\
Chinese Academy of Sciences, 80 Nandan Road, Shanghai 200030, China\\
$^{2}$ University of Chinese Academy of Sciences, 19A Yuquan Road, Beijing 100049, China\\
$^{3}$ Department of Astrophysical Sciences, Princeton University, Princeton, NJ 08544, USA\\
$^{4}$ Department of Astronomy, Columbia University, NYC, NY 10027, USA\\
$^{5}$ Department of Physics, Chongqing University, Chongqing 400044, China }

\pagerange{\pageref{firstpage}--\pageref{lastpage}}

\label{firstpage}

\maketitle

\begin{abstract}
\\
We perform time-dependent, two-dimensional, hydrodynamical, numerical simulations to study the dynamics of a slowly rotating accretion flow from sub-pc to pc scales under the irradiation from the central AGN. Compared to previous work, we improve the calculation of the radiative force due to X-rays. More importantly, in addition to radiative pressure and radiative heating/cooling directly from the central AGN, in the momentum equation we also include the force due to the scattered and reprocessed photons. We find that the accretion flow properties change significantly due to this ``re-radiation'' effect. The  inflow rate at the inner boundary is reduced, while the outflow rate at the outer boundary is enhanced by about one order of magnitude. This effect is more significant when the density at the outer boundary is higher. The properties of outflows such as velocity, momentum and energy fluxes, and the ratio of outflow rate and the accretion rate, are calculated.  We find that the efficiency of transferring the radiation power into the kinetic power of outflow is typically $10^{-3}$, far below the value of $\sim 0.05$ which is assumed in some cosmological simulations. The effect of the temperature of the gas at the outer boundary ($T_0$) is investigated. When $T_0$ is high, the emitted luminosity of the accretion flow oscillates. This is because in this case the gas around the Bondi radius can be more easily heated to be above the virial temperature due to its high internal energy. Another question we hope to address by this work is the so-called ``sub-Eddington'' puzzle. That is, observations show that the luminosity of almost all AGNs are sub-Eddington, while theoretically the luminosity of an accretion flow can easily be super-Eddington. We find that even when the re-radiation effect is included and outflow does become much stronger, the luminosity, while reduced, can still be super-Eddington. Other observational implications and some caveats of our calculations are discussed.
\end{abstract}

\begin{keywords}
accretion, accretion discs - hydrodynamics - methods: numerical - galaxies: active - galaxies: nuclei
\end{keywords}

\section{Introduction}

Active galactic nuclei (AGNs) are thought to play an important role in the processes of galaxy formation and evolution. The most important evidence is perhaps the remarkable
correlations between host galaxy properties and the mass of the supermassive black holes (e.g., Magorrian et al. 1998; Ferrarese \& Merritt 2000; Gebhardt et al. 2000; Graham et al. 2001; Kormendy et al. 2009) observed in the past decade.
Many theoretical studies  have also confirmed that AGN feedback can affect the ambient environment effectively (e.g., Silk \& Rees 1998; Ciotti \& Ostriker 1997, 2001, 2007; Ciotti, Ostriker \& Proga 2009; King 2003; Merloni et al. 2004; Springel, Di Matteo \& Hernquist 2005a; Sazonov et al. 2005; Di Matteo, Springel, \& Hernquist 2005; Hopkins et al. 2005; Murray, Quataert \& Tompson 2005; Somerville et al. 2008; Ostriker et al. 2010).

The AGN feedback can be in the forms of radiative (e.g., Ciotti \& Ostriker 1997, 2001, 2007, Ciotti et al. 2009) or mechanical energy input (e.g., Springel, Di Matteo \& Hernquist 2005b; Ostriker et al. 2010). In terms of the radiative feedback, both momentum feedback and energy feedback work, i.e., the ambient gas could be accelerated by radiation pressure and heated by radiative heating. Mechanical feedback is via wind/outflow. For example, in the context of hot accretion flow (e.g., advection-dominated accretion flow), Yuan, Xie \& Ostriker (2009; see also references therein) studied the effect of global Compton scattering, namely the scattering between the photons produced at the innermost region of the ADAF and the gas at large radius. They found that when the luminosity is higher than $\sim 2\%L_{\rm Edd}$, the heating is so strong that the flow will be heated to be above the virial temperature thus the accretion will be stopped. It is shown that this physical mechanism is able to explain the intermittent activity of some compact young radio sources (Yuan \& Li 2011).

Provided the gas is moderately ionized and can interact with the UV continuum through many UV line transitions, radiation pressure force due to spectral lines (i.e., line force) has been shown to be very efficient at producing winds from accretion disks. This can occur in the small scale of the innermost accretion flow (e.g., Murray et al. 1995; Proga, Stone \& Kallman 2000, hereafter PSK00) and much larger scale from $\sim 0.01$ pc to 10 pc (Proga 2007a; Proga, Ostriker \& Kurosawa 2008; Kurosawa \& Proga 2009, hereafter KP09).  Some highly blueshifted broad absorption line features seen in the UV and optical spectra of AGN can be explained by this mechanism (e.g. Chartas, Brandt \& Gallagher 2003; Crenshaw, Kraemer \& George 2003; Proga \& Kallman 2004; Hamann et al. 2008). However, not all AGN outflows can be explained by line-driven mechanism because of, e.g., the too-low luminosity, too-high ionization state, or both (e.g., Chelouche \& Netzer 2005; Kraemer et al. 2006; see Proga 2007b for a review). This has also been shown to be the case of black hole X-ray binaries which are believed to be the scaled-down version of AGNs (e.g., Miller et al. 2006, 2008; Luketic et al. 2010).

Therefore, other mechanisms are required, such as thermal (e.g., Begelman, de Kool \& Sikora 1991) and, especially, magnetic driving (e.g., Blandford \& Payne 1982; Emmering, Blandford \& Shlosman 1992). Most recently, based on global HD and MHD numerical simulations, Yuan, Bu \& Wu (2012; see also Li, Ostriker \& Sunyaev 2013) show that outflow must exist in hot accretion flows. Moreover, Yuan, Bu \& Wu (2012) proposed a Blandford \& Payne-like mechanism of producing winds. This mechanism should be quite universal since it is a result of magneto-rotational instability (MRI), which is widely believed to exist in accretion flows and is the mechanism of transporting the angular momentum. Previous works on studying the production of MHD winds usually require the pre-existence of large scale poloidal field and treat the disk as the boundary condition. Different from these works, Yuan, Bu \& Wu (2012) simulate the disk and outflow self-consistently and simultaneously, and the pre-existence of the large-scale poloidal magnetic field is not required. However, the simulation was done for hot accretion flows. Thus the next step it remains to be probed explicitly whether it can be equally effective for thin disks, although in principle it is expected to work.

Among the above-mentioned line-driven works, KP09 studied the momentum and energy interaction between the radiation coming from the central AGN and the accretion flow in the region between $\sim 10^{-2}$ pc and $7$ pc. They focus on the line-driven outflow. Compared with previous works, the luminosity of the central AGN is not fixed, but self-consistently determined by evaluating the mass accretion rate at the inner boundary. They found that outflow can be driven efficiently from the inflow. When the temperature of the gas at the outer boundary is high, $T_0\ga 2\times 10^7{\rm K}$, a strong correlation between the mass outflow rate ($\dot M_{\rm out}$) at the outer boundary and the luminosity ($L$) exists, $\dot{M}_{\rm out}\propto L^q$ with $q\sim 2$. When $T_0=2\times 10^6{\rm K}$ {\it and} when $L\ga L_{\rm Edd}$, the correlation becomes flatter, $0\la q\la 0.2$. Another interesting result is, they found that when the density at the outer boundary is high, the accretion luminosity of the central AGN can well exceed the  Eddington value. The super-Eddington accretion mainly proceeds in the equatorial plane and is possible because the radiation flux from the disc is significantly reduced in the equatorial direction due to the geometrical foreshortening effect (Fukue 2000; Watarai et al. 2005).

In almost all previous works on radiatively driven outflow including KP09, they properly consider the forces due to electron scattering and UV line processes due to the ``first-order'' radiation coming from the central AGNs.  However, the effect of the locally produced photons via scattering, the bremsstrahlung, and line radiation, are all neglected. Thus when radiation is absorbed and then re-emitted the dynamical effects of the latter are ignored, essentially violating energy conservation. In principle these photons could play an important role. For example, they can produce an additional force in the vertical direction and thus puff the disk up. This may make the interaction between the radiation and the gas significantly stronger. In the present paper we want to investigate this ``re-radiation'' effect.

Another motivation to consider this effect in the present work is that we hope to address the so-called following ``sub-Eddington'' problem. Observations to a large sample of AGNs with 407 sources show that the distribution of their Eddington ratio ($\equiv L_{\rm bol}/L_{\rm Edd}$) is well described by lognormal, with a peak at $L_{\rm bol}/L_{\rm Edd}\approx 1/4$ and a small dispersion of 0.3 dex (Kollmeier et al. 2006). In other words, almost all AGNs are radiating below $L_{\rm Edd}$. Later  Steinhardt \& Elvis (2010) extended this study to a much larger sample consisting of 62185 quasars from the Sloan Digital Sky Survey and they reached the similar conclusion (but see Kelly \& Shen 2013 for a different opinion). On the other hand, it has long been known from the black hole accretion theory that the rotating accretion flow can well radiate above $L_{\rm Edd}$. Such an accretion flow is called slim disk or optically thick advection-dominated accretion flow (Abramowicz et al. 1988). This analytical result has later been confirmed by radiative hydrodynamical numerical simulations (Ohsuga et al. 2005).

It has been argued that the galactic-scale fueling events are likely to have a broad mass distribution so this should not be the reason for the sub-Eddington luminosity of AGNs. One possible way to solve this puzzle is that, the accretion rates are determined by the self-regulation of the radiative feedback. In other words, the strong radiation from the inner accretion flow may drive a strong outflow thus reducing the accretion rate to below the Eddington rate ($\dot{M}_{\rm Edd}\equiv 10L_{\rm Edd}/c^2$). This should not be effective in the innermost region close to the black hole, since the interaction between radiation and accretion flow has been properly considered in Ohsuga et al. (2005). At larger scale, however, the simulations by KP09 still find super-Eddington accretion even though the radiative feedback is included in their simulations. In this paper, we include the re-radiation effect to do the calculation again, to see whether the ``sub-Eddington'' puzzle can be solved. It is vitally important that the effects of radiation feedback at the Bondi radius be correctly included, since it is in this region that the rate of inflow is determined.

In this paper, following the approach of KP09, we perform two-dimensional hydrodynamical (HD) time-dependent simulations of accretion flow irradiated by the angle dependent UV flux from an accretion disk and isotropic X-ray flux from a corona. The models include radiation forces due to electron scattering and line processes from both the first-order radiation and the re-radiation, and radiative cooling/heating. Special attention is paid to the re-radiation effect. In \S2, we describe some basic aspects of our model, especially  the radiative force due to re-radiation. The numerical simulation method and the initial and boundary conditions are also described in this section. The results of the simulations are given in \S3, and \S4 provides some discussion. Finally, \S5 is devoted to a summary.

\section{Model}

The basic scenario of the model, together with most of the formula for calculating the radiative cooling and heating, are the same with those presented in KP09 (see also PSK00). The readers are referred to that paper for details. However, we do make some changes compared to KP09. In addition to the inclusion of reradiation force mentioned in \S1, we also improve, we believe, the calculation of the radiation force due to X-ray.

\subsection{Hydrodynamics of the accretion flow}
Our aim is to study the dynamics of the accretion flow between an inner radius $r_i$  and outer radius $r_o$. In addition to the gravitational force, this gas will inevitably be irradiated by the radiation emitted from the central AGN. The radiation will heat and push the gas by Compton scattering and line absorption, if the gas is not fully ionized. The dynamics of the accretion flow is described by the following set of equations.
\begin{equation}
   \frac{D\rho}{Dt} + \rho \nabla \cdot {\bf v} = 0,
   \ \ \ \ \ \ \ \ \
\end{equation}
\begin{equation}
   \rho \frac{D{\bf v}}{Dt} = - \nabla P + \rho {\bf g}
 + \rho {\bf F}^{\rm rad},
\end{equation}
\begin{equation}
   \rho \frac{D}{Dt}\left(\frac{e}{ \rho}\right) =
   -p \nabla \cdot {\bf v} + \rho \cal{L}  ,
\end{equation}
where $\rho, e, p$, and ${\mathbf v}$ are the mass density, internal energy density, gas pressure, and velocity, respectively; $\rho\cal{L}$ describes the radiative cooling and heating, simply as the net heating rate (see PSK00);
${\bf g}$ is the gravitational acceleration of the central object;
${\bf F}^{\rm rad}$ (see equation \ref{eq:frad}) is the total radiation force per unit mass. We adopt an adiabatic equation of state $p~=~(\gamma-1)e$, and consider models with the adiabatic index $\gamma=5/3$, with, of course, allowance for heating and cooling via the $\cal{L}$ term in equation (3). The implementations of ${\bf F}^{\rm rad}$ and $\rho\cal{L}$ are described as follows.

\subsection{The central AGN}

The central AGN is described by a super-massive black hole with mass $M_{\rm BH}$ surrounded by an accretion flow. The nature of the accretion flow in a luminous AGN is still an unsolved problem. Usually authors assume that it is described by a standard thin disk. However, a thin disk alone can't explain the origin of hard X-ray emission widely detected in AGNs. For simplicity, we assume that the accretion flow is a combination of a standard thin disk (Shakura \& Sunyaev 1973) plus a spherical corona. They are responsible for the optical/UV and X-ray emissions respectively. The radiation from the corona is assumed to be isotropic.

In the point-source approximation limit, the radiation flux from the
central X-ray corona can be written as
\begin{equation}
    \mathcal{F}_{*}=\frac{f_* L_{acc}\exp{(-\tau_X)}}{4\pi r^{2}},
    \label{eq:corona-flux}
  \end{equation}
where $f_*$ is the fraction of the total luminosity ($L_{\rm acc}$) in X-ray. We adopt $f_*=0.05$ throughout the paper. $\tau_{\rm X}$ is the X-ray optical depth in the radial direction,
  \begin{equation}\label{eq:taux}
  \tau_{\rm X}= \int_0^{r} \rho \kappa_{\rm X}~dr,
  \end{equation}
where $\kappa_X$ is the X-ray opacity. We assume that the attenuation is dominated by Thomson scattering, i.e., $\kappa_X=0.4~{\rm cm^2~g^{-1}}$. However, the scattering opacity must be over-estimated. This is because unlike true absorption, scattering merely re-directs the photons. The scattered X-ray photons must be replenished by photons scattered from other propagation lines. So another extreme is to take $\kappa_X=0$. In the present paper, we also test this case for comparison (see \S4.2). The real opacity should be a certain value between these limits.  The radiation from the standard thin disk $\mathcal{F}_d\propto \left|\cos\theta \right|$. Consequently, the disc radiation flux at a distance $r$ from the center can be written as
  \begin{equation}
    \mathcal{F}_d=2\,\left|\cos\theta\right|\,\frac{f_d  L_{acc} \exp{(-\tau_{UV})}}
    {4\pi r^{2}}
    \label{eq:disk-flux}
  \end{equation}
where $f_d = 1-f_*$ is the fraction of the total luminosity in the disc emission, $\tau_{UV}$ is the UV optical depth and we assume $\tau_{UV}\sim 0$~\footnote{We adopt this approximation since the formula for calculating the line force multiplier only applies to UV optically thin case (cf. the related discussion in KP09). For the same reason, the $\tau_{UV}$ term does not appear in equation (\ref{eq:radiativeforce}). We did some test calculations by including $\tau_{UV}$ and found that the results change only slightly. But note that the gas can still get momentum and energy from the UV radiation although the radiation does not loss its momentum and energy. In this sense, the conservation of momentum and energy is not guaranteed.}. Note that the ``$\left|\cos\theta\right|$'' assumption is potentially a strong assumption, since it suppresses the radiation force close the equatorial plane if we include re-radiation. We therefore have done several test simulations by replacing ``$\left|\cos\theta\right|$'' with ``$(\tau_{perp} + \left|\cos\theta\right|)/(2\tau_{perp} + 1)$''. Here $\tau_{prep}$ is the scattering optical depth  above the disk plane. We find that this effect does not influence our results significantly.

In our simulation, at each time step, the accretion luminosity will be calculated self-consistently based on the mass inflow rate at the inner boundary $r_i$. It is given by
\begin{equation}
L_{acc}\left(t\right)=\eta \dot{M}_{\mathrm{a}}(t) c^2,
\end{equation}
where $\dot{M}_{\mathrm{a}}\left(t\right)$ is the mass inflow rate at $r_i$ and at a given time $t$, $\eta$ is the radiative efficiency. We adopt $\eta=1/12$ in this paper. Following KP09, the mass accretion rate at a given time $t$ is actually time-averaged over some time interval to reduce the fluctuation level. Obviously, a time lag $\tau$ is expected between the change of this accretion rate to the disk and the change of the luminosity from the central AGN $L_{\rm acc}$. Since in the present paper the accretion flow we consider has small angular momentum, a small disk with a radius of $R_d$ will be formed within the inner boundary of our simulation. In this case, the time lag is roughly equal to the accretion timescale at the outer boundary of this small disk. If the small disk can be described by a standard thin disk, the lag time $\tau$ can be approximated as
$t_{\mathrm{acc}}\approx R_d/v_r=R_d/(\alpha c_s H/R_d)$; if it is a slim disk, the time lag should be $t_{\mathrm{acc}}\approx R_d/v_r \sim R_d/(\alpha v_k)$. Here $\alpha$ is the viscosity parameter, and we set $\alpha=0.03$ in this paper. The black hole mass is assumed to be constant. The value of $\tau$ is not important for our aim.

One caveat here is that we assume the accretion rate onto the black hole is equal to the inflow rate at $r_i$. This may not be true, however. Almost all global numerical simulations to hot accretion flows show that the accretion rate should decrease with decreasing radius (see Yuan, Wu \& Bu 2012 for a review). The reason is identified to be because of continuous mass loss in outflow produced by an MHD process similar to the Blandford-Payne mechanism (Yuan, Bu \& Wu 2012). This mechanism in principle also works in the  case of a cold thin disk. Actually, the initial global radiation MHD numerical simulations to thin disk by Ohsuga et al. (2009; see also Ohsuga \& Mineshige 2011) do show significant outflow, although the exact mechanism remains to be probed. Since the outflow in the context of a thin disk is still an open question, in the present paper we simply assume that the accretion rate is constant within $r_i$ and thus our computed luminosity is an upper bound on the true luminosity for given outer boundary conditions.

\subsection{Heating/cooling and radiative force of the gas}

The interaction between the radiation and the gas and the radiative processes we consider includes Compton heating/cooling, X-ray photoionization and
recombination, bremsstrahlung, and line cooling (cf. PSK00; see also Blondin 1994). The net cooling
rate depends on the photoionization parameter $\xi$ and the characteristic
temperature (or radiation temperature) of the radiation from the central AGN ($T_X$).
The gas photoionization state is determined by the photoionization parameter
\begin{equation}
\xi = \frac{4 \pi {\cal F}_*}{n}={f_* L_{acc} \exp{(-\tau_X)}}/{n r^2} ,
\end{equation}
where $n=\rho/\mu m_p$ is the number density of the local gas located at distance $r$ from the central AGN, $\mu$ is the mean molecular weight and is fixed to be 1. Only X-ray radiation ionizes the gas. The sum of photoionization heating-recombination cooling rate ($n^2G_X$) and Compton heating/cooling rate ($n^2G_{Comp}$) is \begin{equation} \dot E_X=n^2(G_X+G_{Comp}), \label{xrayheating}\end{equation} with
\begin{equation}
\label{eq:Gx}
G_{X}~=~1.5~\times10^{-21}~\xi^{1/4}~T^{-1/2}(1-T/T_{\rm X})
~\rm~erg~\rm cm^{3}~s^{-1},
\end{equation}
\begin{equation}
\label{eq:Gcomp}
G_{Comp}~=8.9\times10^{-36}~\xi~(T_{\rm X}-T)
~\rm~erg~\rm cm^{3}~s^{-1}.
\end{equation}
Here $T_X\sim8\times10^7 {\rm K}$ (Sazonov et al. 2004) is the ``characteristic temperature'' of the X-ray radiation and it is $4$ times larger than the Compton temperature (refer to PSK00 and Proga 2007a). For a low-luminosity AGN, $T_X$ will be much higher, $\sim 3 \times 10^9$K (Yuan, Xie \& Ostriker 2009).

\subsection{Radiative force} \label{sec:rforce}
\begin{table*}
\caption{Summary of models with different parameters.}
\label{tab:model-summary}

\begin{tabular}{lcccccccc}
\hline
Model  & Model &     $r_i$   &  $\rho_0$  &  $T_0$  & Re-radiation& $L_{acc}$ &  $\dot M_{out}(r_o)$  & $\eta_w$ \tabularnewline
Number &       &$(r_{\ast})$ &$(10^{-21} {\rm g~cm^{-3}})$& $(10^{6}{{\rm K}})$ &  & $(L_{Edd})$ & $(10^{25} {\rm g~s^{-1}})$ & \tabularnewline
(1)    &  (2)  &    (3)      &     (4)    &  (5)    &   (6)       &   (7)     &   (8)    & (9) \tabularnewline
\hline
1      & K5a   &500  &10   & 2  & no  & 1.19(0.19)     &   15.31(9.12)  & 0.76(9.12)\tabularnewline
2      & K6a   &500  &20   & 2  & no  & 1.85(0.43)     &   25.61(24.29) & 0.84(0.92)\tabularnewline
3      & K6c   &1250 &20   & 2  & no  & 1.72(0.14)     &   29.05(17.12) & 0.98(0.57)\tabularnewline
4      & K6d   &2550 &20   & 2  & no  & 1.71(0.23)     &   31.17(13.22) & 1.10(0.53)\tabularnewline
5      & K7a   &500  &50   & 2  & no  & 3.97(1.10)     &   21.54(27.46) & 0.35(0.58)\tabularnewline
6      & K7b   &625  &50   & 2  & no  & 3.69(0.78)     &   20.80(17.05) & 0.33(0.24)\tabularnewline
7      & K7d   &2500 &50   & 2  & no  & 3.14(1.03)     &   49.82(42.21) & 1.03(0.95)\tabularnewline
8      & K8a   &500  &100  & 2  & no  & 5.74(0.70)     &   21.52(14.41) & 0.22(0.13)\tabularnewline
9      & K8d   &2500 &100  & 2  & no  & 5.70(1.75)     &   34.24(50.46) & 0.35(0.51)\tabularnewline
$\cdots$ & $\cdots$ & $\cdots$ & $\cdots$ & $\cdots$ & $\cdots$ & $\cdots$ & $\cdots$ & $\cdots$\tabularnewline
10     & R5a   &500  &10   & 2  & yes & 0.77(0.04)     &  37.95(8.71)  & 2.88(0.70)\tabularnewline
11     & R5b   &625  &10   & 2  & yes & 0.94(0.04)     &  31.53(5.07)  & 1.96(0.34)\tabularnewline
12     & R5c   &1250 &10   & 2  & yes & 1.10(0.10)     &  24.28(10.67) & 1.28(0.55)\tabularnewline
13     & R5d   &2500 &10   & 2  & yes & 1.12(0.11)     &  26.88(8.05)  & 1.40(0.45)\tabularnewline
14     & R6a   &500  &20   & 2  & yes & 1.19(0.16)     &  67.30(10.90) & 3.35(0.78)\tabularnewline
15     & R6b   &625  &20   & 2  & yes & 1.28(0.17)     &  65.14(10.73) & 3.02(0.69)\tabularnewline
16     & R6c   &1250 &20   & 2  & yes & 1.48(0.18)     &  39.10(12.13) & 1.54(0.45)\tabularnewline
17     & R6d   &2500 &20   & 2  & yes & 1.60(0.18)     &  44.04(11.45) & 1.61(0.45)\tabularnewline
18     & R7a   &500  &50   & 2  & yes & 2.02(0.02)     &  139.13(5.03) & 3.99(0.14)\tabularnewline
19     & R7b   &625  &50   & 2  & yes & 2.03(0.03)     &  137.86(4.83) & 3.93(0.14) \tabularnewline
20     & R7c   &1250 &50   & 2  & yes & 1.81(0.03)     &  127.04(11.80)& 4.08(0.39) \tabularnewline
21     & R7d   &2500 &50   & 2  & yes & 1.60(0.03)     &  194.29(1.92) & 7.04(0.14) \tabularnewline
22     & R8b   &625  &100  & 2  & yes & 3.14(0.27)     &  202.05(28.40)& 3.75(0.59) \tabularnewline
23     & R8c   &1250 &100  & 2  & yes & 4.56(1.09)     &  36.45(54.69) & 0.49(0.82) \tabularnewline
24     & R8d   &2500 &100  & 2  & yes & 4.14(1.20)     &  126.12(97.44)& 2.14(1.91) \tabularnewline
$\cdots$ & $\cdots$ & $\cdots$ & $\cdots$ & $\cdots$ & $\cdots$ & $\cdots$ & $\cdots$ & $\cdots$\tabularnewline
25     & R7c-X0    &1250 & 50  & 2   & yes & 1.76(0.22)     &  120.61(9.67)  & 4.03(0.58) \tabularnewline
26     & R7c-lT    &1250 & 50  & 0.5 & yes & 1.61(0.27)     &  28.59(11.05)  & 1.08(0.50) \tabularnewline
27     & R7c-lTX0  &1250 & 50  & 0.5 & yes & 1.19(0.01)     &  52.17(3.28)   & 2.54(0.16) \tabularnewline
28     & R7c-hT    &1250 & 50  & 6   & yes & 2.94(0.04)     &  245.72(10.33) & 4.85(0.22) \tabularnewline
29     & R7c-hTX0  &1250 & 50  & 6   & yes & 2.26(0.46)     &  338.04(29.36) & 8.98(1.85) \tabularnewline
30     & R7c-hT2   &1250 & 50  & 10  & yes & 2.02(0.97)     &  461.93(100.00)& 15.43(5.07) \tabularnewline
31     & R7c-hT2X0 &1250 & 50  & 10  & yes & 2.74(0.51)     &  508.50(57.05) & 11.18(3.12) \tabularnewline
\hline
\end{tabular}

\leftline{Note:~Values in brackets in Columns 7, 8 and 9 are the normalized standard deviations $\sigma_n$ of the time series values. The symbol `X0'}
\leftline{in the model names means $\kappa_X=0$ in those models, `lT' means lower temperature $T_0$, `hT' means higher temperature $T_0$.}
\end{table*}

Compared to KP09, in our present work we make two changes when calculating radiative forces. The radiative forces considered in KP09 are due to Compton scattering and line processes. The former is the only available radiative force if the gas is fully ionized. Obviously, this force is important only when the luminosity from the central AGN approaches the Eddington value. However, the force can be significantly enhanced if the gas is moderately ionized, since the opacity is much higher than scattering opacity due to the bound-bound and bound-free transitions. So we should include the line force. For this force, following KP09, we assume that only optical/UV photons have contributions, but neglect the possible force line due to some metal line in the soft X-ray band.
Regarding the sources of photons, as we have emphasized in \S1, in addition to the radiation from the central AGN, the ``local'' photons from the local radiative processes and scattered photons originally from the central AGNs are also included. Both of them contribute to the radiative force,
\begin{equation} \label{eq:frad}
{\bf F}^{rad}={\bf F}_r^{rad,c}+{\bf F}_z^{rad,re}
\end{equation}
The inclusion of the force due to the ``re-radiation'' process (${\bf F}_z^{rad,re}$) is the first change we make compared to KP09. It turns out that the inclusion of this effect does significantly change the results, as we will state in the following section.

For X-ray photons, KP09 only consider the force due to Thomson scattering. However, if the gas is not fully ionized as in our present case, absorption processes such as photoionization should also contribute to the force. We characterize this force by $\dot{E}_X/(\rho c)$\footnote{We note that this simplification will underestimate the force, since in principle we should use only X-ray heating rate. We find that this approximation is feasible because the force is in general dominated by the line force.}. The inclusion of this force is the second change we make compared to KP09.

Based on the above considerations, the total radiative force due to the central AGN photons then can be written as:
\begin{eqnarray}\label{eq:radiativeforce}
{\bf F}_r^{rad,c}~(r,\theta) = \left[\frac{\dot E_X}{\rho c}+\frac{2\kappa_{es}}{c}\frac{L_{acc}}{4\pi r^2}(1+M)\left|\cos\theta\right|f_{d}\right] \cdot \hat{\bf r}.
\end{eqnarray}
Here $\kappa_{es}=0.4~{\rm cm^2~g^{-1}}$ is the scattering coefficient for free electrons, and $M$ is the force multiplier (Castor et al. 1975) -- the numerical factor which parameterizes by how much spectral lines increase the scattering coefficient (see PSK00 for details). We have calculated the dependence of the multiplier $M$ on the temperature $T$ and the ionization parameter $\xi$. We found that the multiplier $M$ could be up to the order of $10^3 \sim 10^{4}$, providing that the gas is moderately ionized. It drops rapidly when temperature $T$ or the ionization parameter $\xi$ is larger than some critical value (eg., $\xi>10$, or $T>10^5$). Specifically, it is safe to ignore the line force ($\lesssim 1\%$ of electron scattering force) when $\xi>10^3$ or $T>2\times10^6$, since then the gas is almost fully ionized.

Now we calculate the second term in equation (\ref{eq:frad}), i.e, the re-radiation force. Exact treatment of this force requires the full radiative transfer calculation which is beyond the scope of the present paper. Because the accretion flow is rotating (see below), the flow has a disk-like shape. In this case, for simplicity we can adopt the assumption of plane-parallel approximation. The ``re-radiation'' photons would eventually escape from the accreting flow, and produce a net force in the vertical direction. Consider a rectangular box located within the disk, with the bottom of the box overlapped with the equatorial plane of the accretion disk while the upper side at height $z$. ``Re-radiation'' photons are steadily produced within this box and then escape. Given the plane-parallel approximation and symmetry, the net radiative flux due to re-radiation (and thus the force) should be only in the upper side of the box. Applying the Gauss theorem, we can easily calculate the vertical radiative force,
\begin{equation}\label{eq:reradiation}
{\bf F}_z^{rad,re}~(r,\theta)=\frac{\kappa_{es}}{c}\int^z_0 (S_{c}+n^2L_{brem}+n^2L_{line}) dz \cdot \hat{\bf z},
\end{equation}
where
\begin{equation}\label{eq:scv}
S_{c}=
 \rho \kappa_{es} \cdot \frac{L_{acc}}{4\pi r^2}\left[f_\ast\exp{(-\tau_X)}
  +2\left|\cos\theta\right|f_d \right]
\end{equation}
is the source term due to the first-order scattered photons of the radiation from the central AGN; while
\begin{equation} \label{eq:lbrem}
L_{\rm brem}=3.3\times10^{-27}T^{1/2}{\rm erg~cm^{3}~s^{-1}}
\end{equation}
and
\begin{equation} \label{eq:lline}
\begin{aligned}
L_{\rm line}=\delta[&1.7\times 10^{-18}{\rm exp}(-1.3\times10^5/T)\xi^{-1}T^{-1/2}\\
&+10^{-24}]{\rm erg~cm^{3}~s^{-1}}
\end{aligned}
\end{equation}
are the rate of bremsstrahlung and line cooling (cf. PSK00), respectively. The parameter $\delta$ is introduced to control line cooling, $\delta=1$ represents optically thin cooling, and $\delta<1$ represents optically thick cooling. We set $\delta=1$ throughout this paper. In general, the line cooling dominates over the other cooling processes. The formulae above only show the re-radiation force caused by electron scattering while the line force is ignored. So the effect of ``re-radiation'' calculated in this paper should be regarded as a lower limit in this sense. We did include the line force in some models and found that the change of the result is not very significant. This is perhaps because the re-radiation force monotonically increases with height, and at a large height, the gas temperature becomes high so the line force is not important. One thing to note is that all our treatment is approximate; a full radiative transfer calculation is required in the future.

\subsection{Simulation setup}
We solve the dynamical equations (1-3) using the code {\it ZEUS-MP} (Stone \& Norman 1992a; Hayes et al. 2006). The simulations are performed in spherical polar coordinates
$(r,\theta,\phi)$ assuming axial symmetry about the disc rotational axis ($\theta=0^\circ$) and in two-dimensions. Our computational domain is defined to occupy
the angular range $0^{\circ} \leq \theta \leq 90^{\circ}$ and the radial range
$r_{\rm i}\leq r \leq \ r_{\rm o}$. In KP09, both $r_{\rm i}$ and $r_{\rm o}$ are fixed, i.e, they are same in all models. In the present work, we want to check the convergence for various values of $r_{\rm i}$, so we set various $r_{\rm i}$ in different models but adopt $r_{\rm o}=2.5\times10^5r_\ast$,
where $r_\ast=6 r_{g}$ is the innermost stable circular orbit (ISCO hereafter) of a
Schwarzschild BH and $r_{g}\equiv GM_{\rm BH}/c^2$. We set the BH mass, $M_{\rm BH}=10^8\MSUN$. Our standard numerical resolution in the $r$ direction consists of 144
zones with the zone size ratio, $dr_{i+1}/dr_{i}=1.04$ and consists
of 64 zones in the $\theta$ direction with $d\theta_{j+1}/d\theta_{j} =1.0$.
Gridding in this manner ensures good spatial resolution close to
the inner boundary.

\begin{figure*}
\begin{center}
\includegraphics[width=15cm]{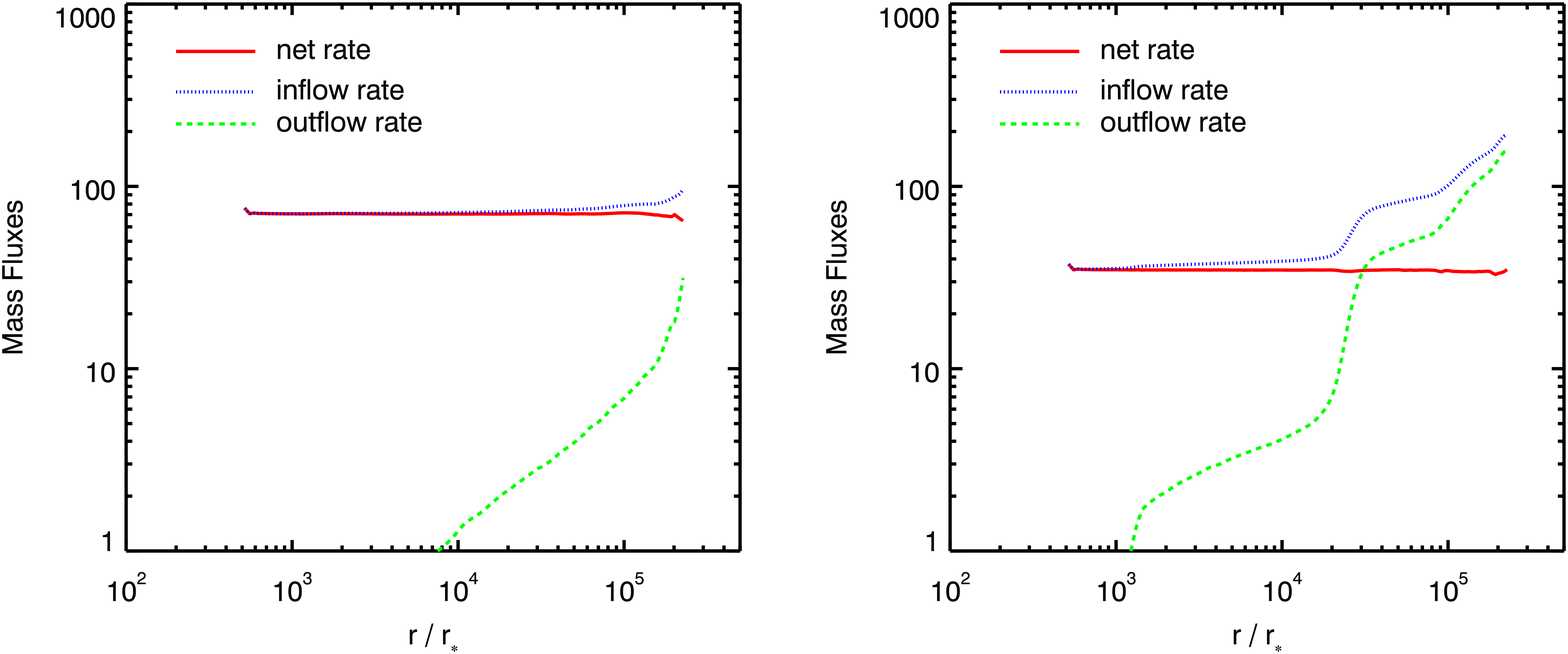}
\includegraphics[width=15cm]{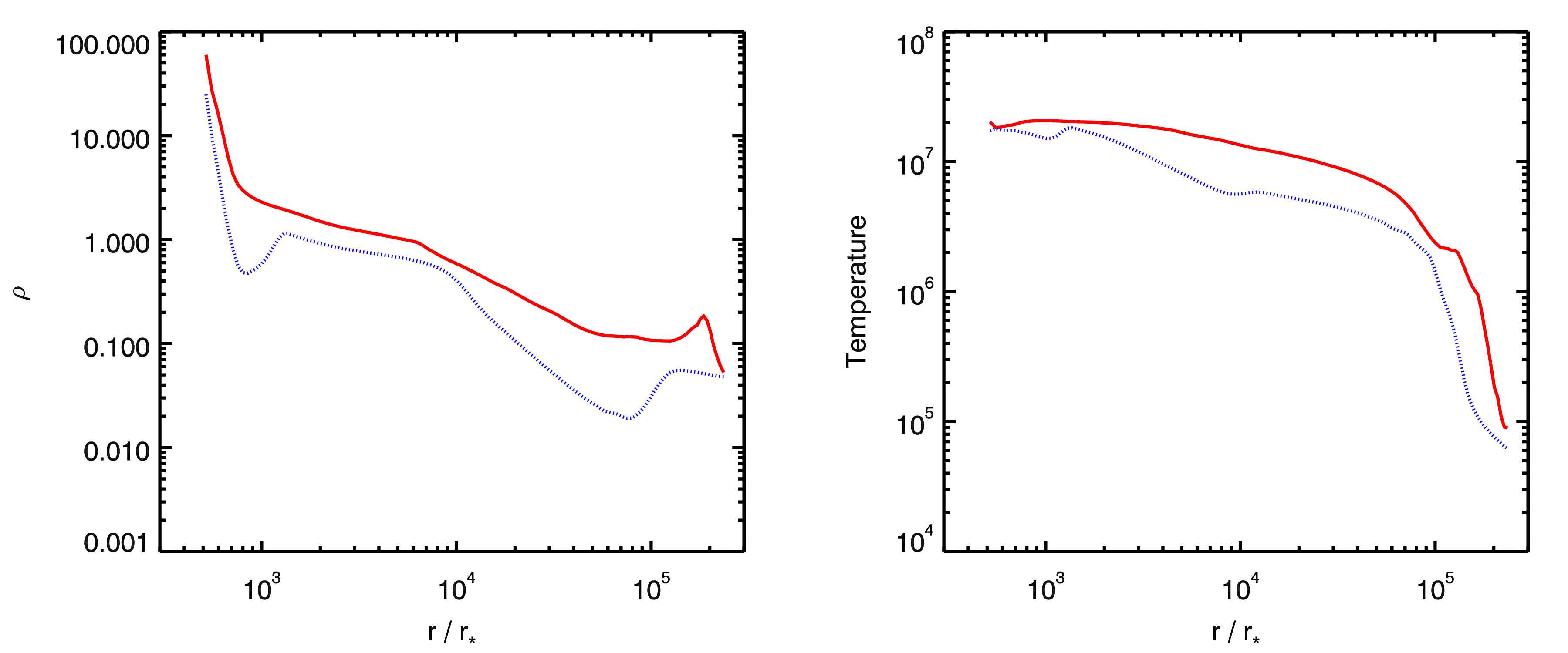}
\end{center}
\caption{Comparison between models Model K7a (without considering re-radiation; top left panel and the red solid line in the bottom panels) and Model R7a (with re-radiation included; top right panel and the blue dotted line in the bottom panels). {\it Top panel}~---~The radial profiles of mass inflow rate (blue dotted line), outflow rate (green dashed line), and the net rate (red solid line) in unit of $10^{25} {\rm g~s^{-1}}$. {\it Bottom panel}~---~The radial profiles of gas density ({\it left panel}, in unit of $\rm 10^{-18}g~cm^{-3}$) and temperature ({\it right panel}, in unit of K) averaged over three grids above the equatorial plane. All the lines are time-averaged.}
\label{fig:compare-7}
\end{figure*}

For the initial condition, following KP09, we set the density and temperature of gas uniformly , i.e., $\rho=\rho_0$ and $T=T_0$ everywhere in the computational domain. The initial velocity of the gas is assigned to have the distribution described by the following latitude-dependent angular momentum at the outer boundary.
\begin{equation}\label{eq:angm}
l(\theta)=l_0 (1-|\cos\theta|), \ \ \ \  l_0=\sqrt{G M_{BH} r_{cir}}
\end{equation}
where $r_{cir}$ is the ``circularization radius'' on the equatorial plane.
In the present work, we set $r_{cir}=300r_*< r_i$. The boundary conditions are specified in the following way.
We apply an axis-of-symmetry boundary condition at the pole (i.e., $\theta = 0^\circ$) and reflecting boundary conditions at $\theta=90^\circ$.
For the inner and outer radial boundaries, we apply an outflow boundary condition (i.e.,
to extrapolate the flow beyond the boundary, we set values
of variables in the ghost zones equal to the values in the corresponding
active zones, see Stone \& Norman (1992a) for more details).
To represent steady conditions at the outer radial boundary,
during the evolution of each model we apply the constraints that
all hydrodynamical variables except radial velocity $v_r$ (always let it float) in the last zone in the radial direction to be equal to  the initially chosen values, i.e., $\rho=\rho_0$, $T=T_0$, $v_\theta=0$ and $v_\phi=l/(r\sin\theta)$, when $v_r(r_o,\theta)<0$ (inflowing). We allow all hydrodynamical variables to float when $v_r(r_o,\theta)>0$ (outflowing). This approach is to mimic the situation where there is
always gas available for accretion.

\begin{figure*}
\includegraphics[width=15cm]{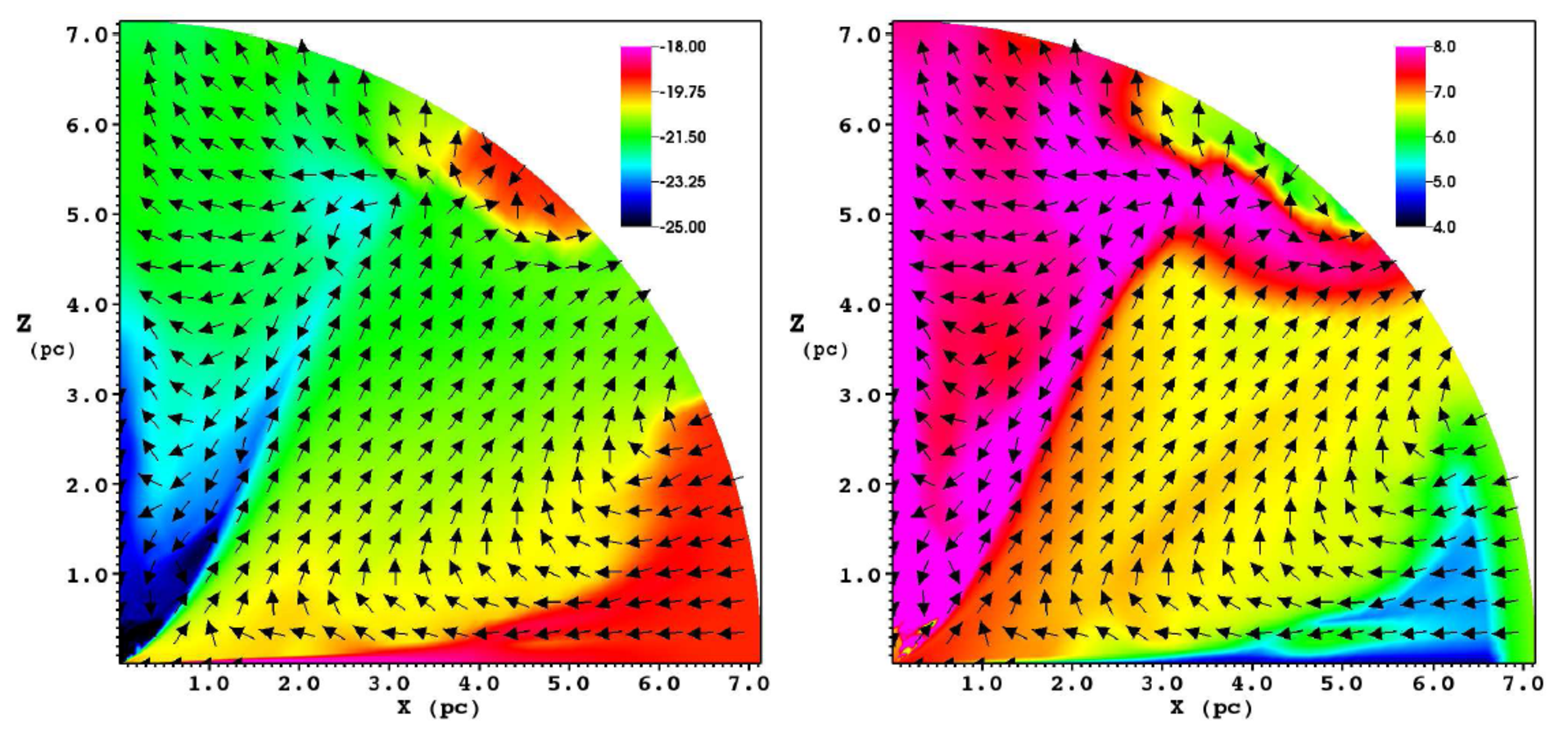}
\includegraphics[width=15cm]{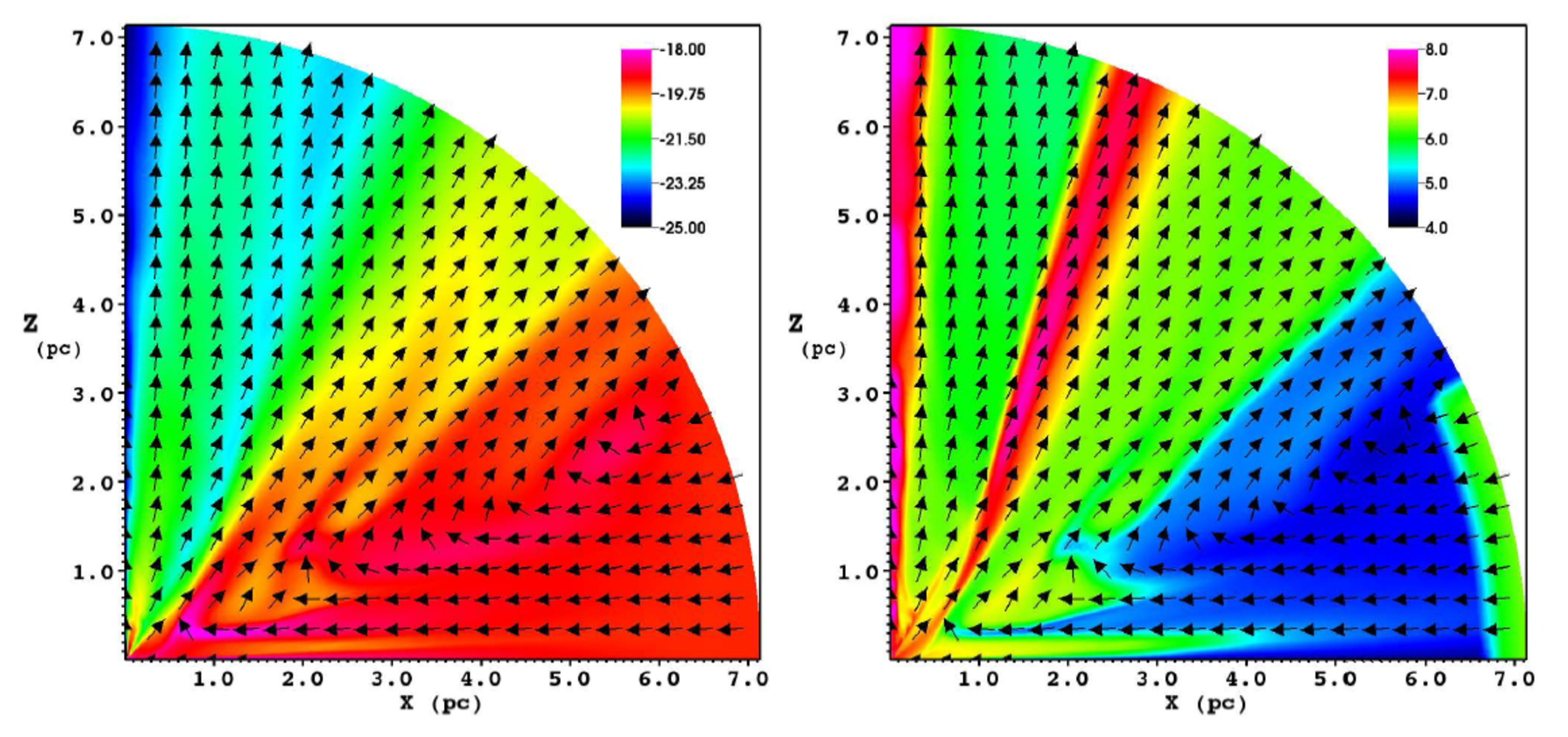}
\caption{Snapshots of contours of logarithmic gas density (left column) and logarithmic temperature (right column) at steady state. The top panel is for Model K7a (without re-radiation), and the  bottom panel is for Model R7a (with re-radiation included).}
\label{fig:contour7}
\end{figure*}

So the free parameters in our model include the inner radius $r_i$ of the computational domain, the density $\rho_0$ and temperature $T_0$ of inflowing matter at the outer boundary. In this paper we focus mainly on the case of $T_0=2\times 10^6 {\rm K}$. But we do also consider other $T_0$ for comparison. Various $r_i$ and $\rho_0$ are adopted. The details of various models are listed in Table \ref{tab:model-summary}. The ``yes/no'' shown in Column 6 indicates whether re-radiation is included in the simulations.

\section{Results}

Some results of various models are given in Table \ref{tab:model-summary}. Especially, Column 7 gives the time-averaged luminosity calculated according to the mass accretion rate at the inner boundary. This may exceed the luminosity as seen at infinity due to absorption of radiation within the flow. Models 1, 2, 5 and 8 are identical to the models 5, 6, 7 and 8 in KP09 except that we use our improved term when calculating X-ray force, i.e., the first term in equation (\ref{eq:radiativeforce}). The last characters in the model name, i.e., ``a'', ``b'', ``c'', and ``d'', correspond to $r_i=500, 625, 1250, 2500$ in units of $r_\ast$, respectively. We define the mass inflow and outflow rates as below:
\begin{equation}\label{eq:inflow}
\dot M_{\rm in}(r)=2\pi r^2\int^{\pi}_0 \rho~{\rm min}(v_r,0)~{\rm sin}(\theta)~d\theta,
\end{equation}
\begin{equation}\label{eq:outflow}
\dot M_{\rm out}(r)=2\pi r^2\int^{\pi}_0 \rho~{\rm max}(v_r,0)~{\rm sin}(\theta)~d\theta.
\end{equation}
The time-averaged mass outflow rate is given in Column 8. The Column 9 gives
\begin{equation}\label{eq:eta_w}
 \eta_w=\frac{\dot M_{out}(r_o)}{\dot M_a},
\end{equation}
which is the ratio of mass outflow rate at the outer boundary and mass accretion rate at the inner boundary. It is important to note that the fraction of the material which is inflowing at the outer boundary which actually accretes on the black hole, defined as $f_{ac}\equiv\dot M_a/\dot M_{in}(r_o)$, is from the mass conservation $f_{ac}=1/(1+\eta_w)$; so that in cases where $\eta_w\gg1$ the net accretion, $f_{ac}$ is small.

\subsection{Effects of re-radiation}

We take models 7 series (i.e., K7* and R7*) as our fiducial models and explore their properties in details. Fig. \ref{fig:compare-7} compares the radial profiles of mass inflow and outflow rates, gas density and temperature between models K7a and R7a.
The former does not consider re-radiation while the latter does. Fig. \ref{fig:contour7} shows the snapshots of contours of gas density and temperature of the two models. From the two figures we can find the following changes after re-radiation is taken into account.
\begin{enumerate}
\item Close to the equatorial plane the gas density decreases when re-radiation is included, as shown by the bottom-left panel of Fig. \ref{fig:compare-7}. However, we can see from Fig. \ref{fig:contour7} that away from the equatorial plane the density becomes higher, i.e., the density scale-height becomes larger. The reason is that the accretion flow expands upward due to the inclusion of the vertical radiative force from re-radiation photons, then the vertically expanded inner edge of the accretion flow shields the radiation from the central AGN, which makes the density higher and the mass inflow rate larger.
\item From Fig. \ref{fig:contour7}, we can see that the temperature becomes lower in most of the region. This is because the increase of density produces stronger radiative cooling. The decrease of temperature close to the equatorial plane is because the compression work induced by accretion becomes weaker due to the inclusion of the vertical force.
\item Perhaps the most important effect of including re-radiation is that outflow becomes nearly one order of magnitude stronger, as clearly shown by the comparison between the top two panels of Fig. \ref{fig:compare-7}.  This is easy to understand.  As we state above, the inclusion of the vertical force due to re-radiation photons makes the density scale-height larger. Since the irradiation force from the central AGN $\propto \cos\theta$, the gas will be more easily blown away if located at higher latitude.
\item The mass inflow rate at the inner boundary decreases by a factor of 2. This is obviously because of the enhanced mass loss via the outflow. It is interesting to note that the mass inflow rate at the outer boundary increases by a factor of 2. The reasons are two-fold. One is that the radiation from the central AGN decreases because of the decrease of the inflow rate at the inner boundary, thus the outward radiation force becomes weaker. Another reason is that, as we stated above, the gas density in most of the region increases. This gas will more effectively shield the gas at the outer boundary from irradiation by the central AGN.
\item The flow morphology also changes with the inclusion of re-radiation. Comparing the top and bottom panels of Fig. \ref{fig:contour7}, we can see that in Model K7a, the inflow falls inward in a relatively small range of polar angle, i.e., $\theta\sim 80^{\circ}-90^{\circ}$. But in Model R7a, i.e, the re-radiation effect is included, the inflow occurs in a relatively larger range, $60^{\circ}\la\theta\la 90^{\circ}$. In addition, it is interesting to  note that above the inflow region, the motion of the gas is turbulent in Model K7a; while in Model R7a, the motion is less turbulent.
\end{enumerate}

KP09 studied the correlation between the mass outflow rate at the outer boundary and the luminosity in units of Eddington luminosity $\Gamma (\equiv L/L_{\rm Edd})$, where $L_{\rm Edd}=4\pi c GM_{\rm BH}/\sigma_e)$ is the Eddington luminosity. They found that when $T_0$ is not too high, $T_0\la 10^8K$, outflows exist only when $\Gamma\ga 0.5$. Specifically, when $T_0=2\times 10^6K$, as in our paper, a strong correlation is found for $0.5\la \Gamma\la 1.5$, i.e., $\dot{M}_{\rm out}\propto \Gamma^q$ with $q=2.3(\pm0.3)$. The correlation index $q$ only becomes slightly smaller when the temperature $T_0$ at the outer boundary is higher. In the case of $T_0=2\times 10^6K$, $\dot{M}_{\rm out}$ becomes almost saturated when $\Gamma\ga1.5$. KP09 explains this result as because the inflow is squeezed to a narrow range of $\theta$ when $\Gamma$ becomes high.

\begin{figure}
\includegraphics[width=7.5cm]{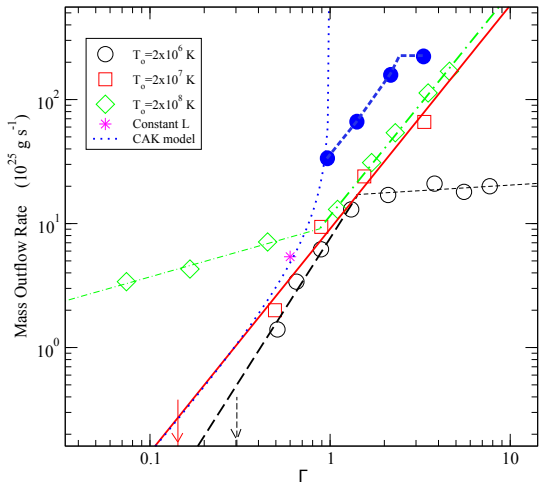}
\caption{{The correlation between the mass outflow rate and Eddington ratio ($\Gamma=L_{acc}/L_{Edd}$). All the lines and dots except the three solid blue ones are taken from KP09. The three solid blue dots are the results when the re-radiation effect is included when $T_0=2\times 10^6{\rm K}$. We can see that the correlation index $q$ almost remains unchanged.}}
\label{fig:newgamma}
\end{figure}

We have studied the effect of re-radiation on the correlation. Fig. \ref{fig:newgamma} shows the result. All the lines and dots except the four solid blue dots are taken from KP09. The four solid blue dots are the results when re-radiation effect is included  for $T_0=2\times 10^6{\rm K}$. We can see that the correlation index $q$ roughly remain the same, and the correlation again becomes almost saturated when $\Gamma\ga 2$. The only change is the normalization: the outflow rate increases by almost one order of magnitude as we have stated above.

At last, as we have mentioned above, the re-radiative force makes the density in most of the region larger. In this case, the attenuation of X-rays also becomes much more significant, thus the photoionization parameter $\xi$ decreases. As we have mentioned in section \ref{sec:rforce}, the line force multiplier usually increases with decreasing $\xi$. In other words, the line force should increase rather than decrease when we consider the re-radiation effect.

\subsection{Convergence with the inner boundaries}

It is necessary to check whether the result converges with various inner boundaries $r_i$. With this aim, we have implemented a series of runs (see Table \ref{tab:model-summary}) by varying the value of $r_i$.

\begin{figure*}
\includegraphics[width=15.cm]{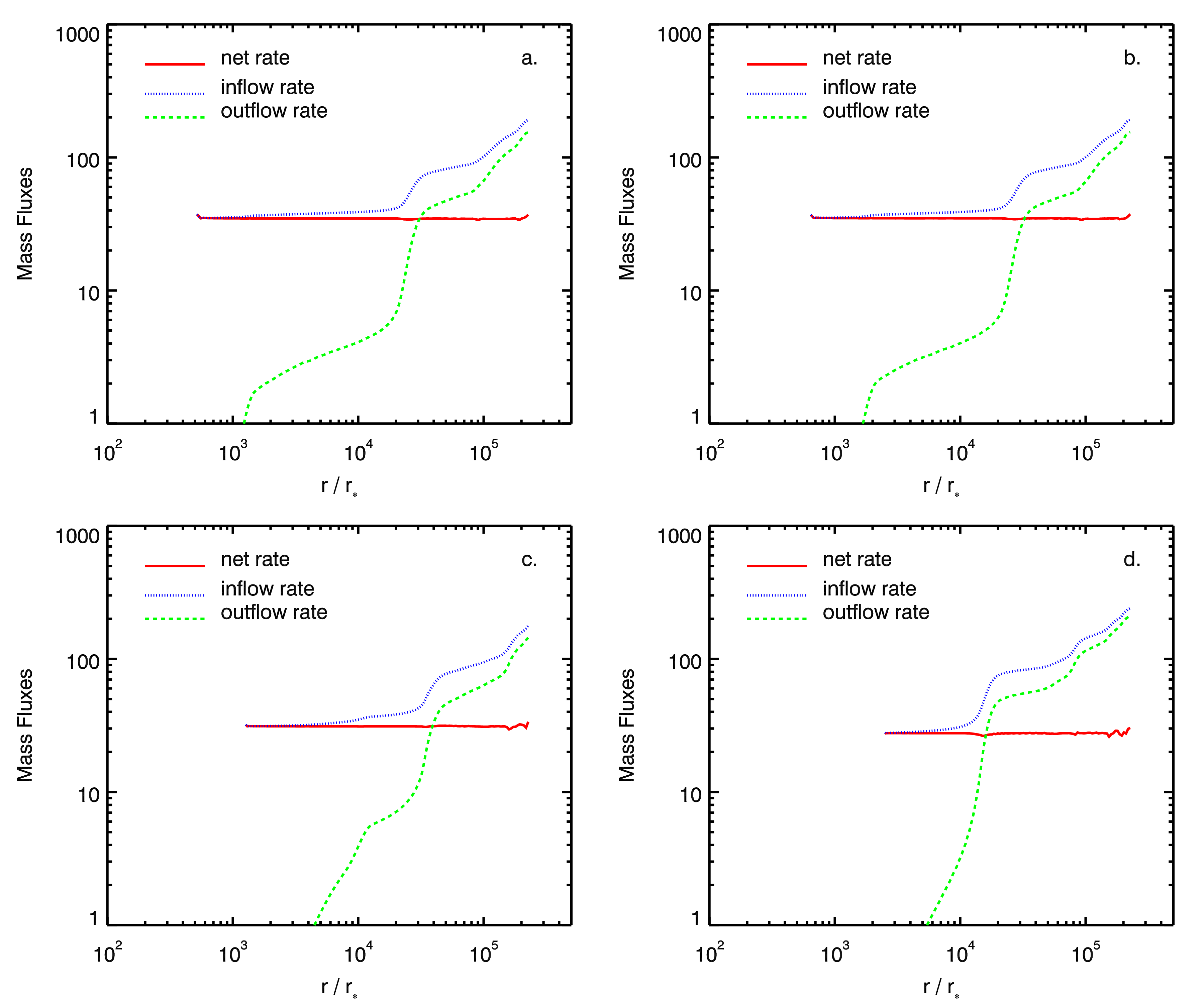}
\includegraphics[width=15.cm]{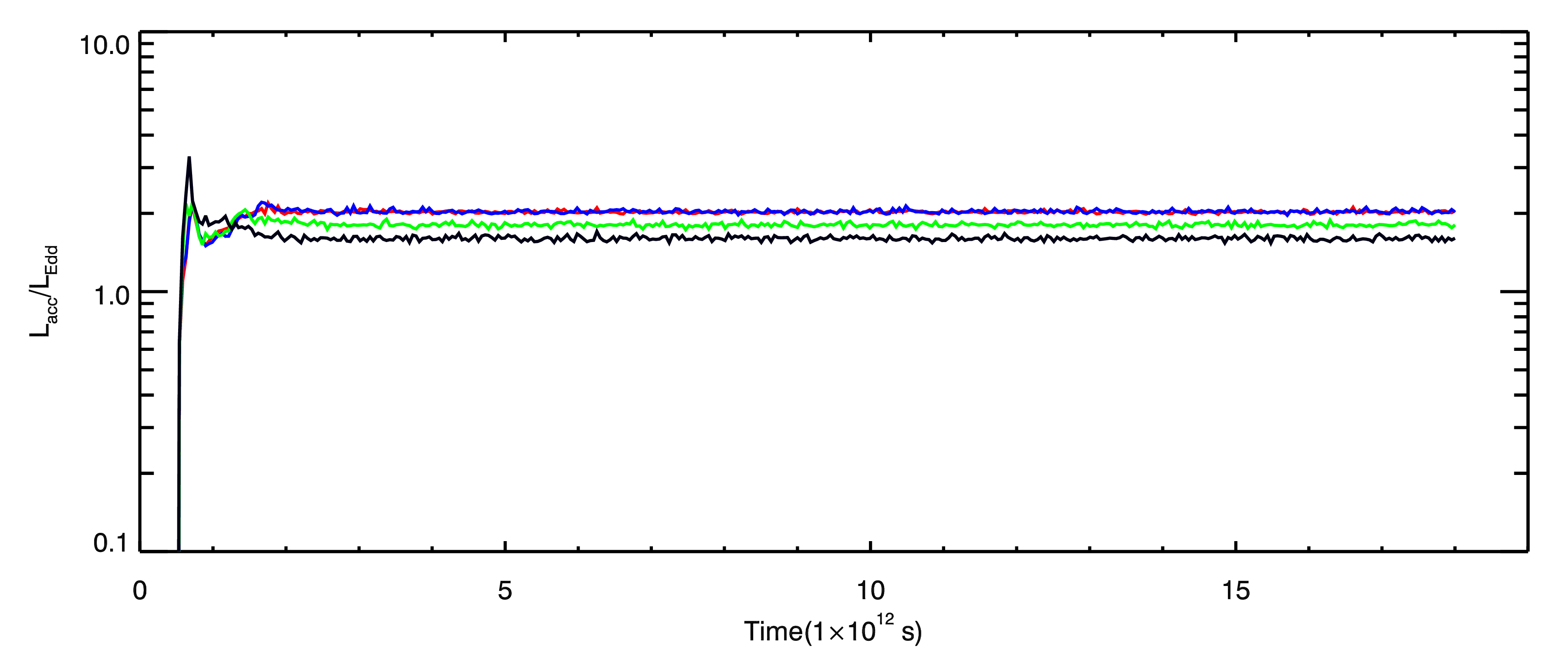}
\caption{Effects of varying the inner boundary $r_i$ for model R7 series. {\it Top and middle panels:} The radial profiles of mass inflow rate, outflow rate, and the net rate in unit of $10^{25} {\rm g~cm}^{-3}$. Models a, b, c, and d correspond to $r_i=500r_\ast, 625r_\ast, 1250r_\ast$, and $2500r_\ast$, respectively. We can see the results are roughly convergent with various $r_i$.  {\it Bottom panel}: Time evolution of luminosity normalized by Eddington luminosity $L_{\rm Edd}$ for model R7 series. The red, blue, green, and black lines correspond to models R7a, R7b, R7c, and R7d, respectively. All the lines in the top and middle panels are time-averaged.}
\label{fig:mfluxall7}
\end{figure*}

Fig. \ref{fig:mfluxall7} shows the simulation results for the model R7 series. The top and middle panels show the radial profiles of inflow, outflow, and net rates for the four models with different $r_i$, while the bottom panel shows the time variation of the luminosity of the models.
The values of the outflow rate of the four models are almost same. Moreover, the radius where the outflow begins to become equal to the net rate is also roughly same, i.e., $\sim (1-4)\times 10^4 r_\ast$. For the inflow rate at the inner boundary, models R7a and R7b are almost same; while
model R7c is only about 10\% smaller than model R7a and model R7d is only about 12\% smaller than model R7c. The deviations are within the acceptable range. Therefore, we think model R7 series are convergent as we vary the radius of the inner boundaries.

We also did the simulations with different $r_i$ for model R5, R6 and R8 series. The results are also listed in Table \ref{tab:model-summary}.
For model R5 series, the largest discrepancy of time-averaged luminosity among the four models is $\sim 36\%$. For model R6 and R8 series, they are $25\%$ and $31\%$, respectively. We found that the results are similar when re-radiation is not included, as shown by Models K6a, K6c, and K6d. The change of the luminosity with $r_i$ is complicated: the luminosity can increase or decrease with increasing $r_i$. Physically, when $r_i$ increases, the attenuation of X-ray flux will become weaker, thus the corresponding radiation force becomes stronger. But on the other hand, when X-ray radiation becomes weaker, the ionization will become weaker and correspondingly the line force stronger. Thus the final result depends on the competition of these two effects. But one relatively robust result is that the outflow mainly originates from $\sim (1-4) \times 10^4 r_\ast$. Another noteworthy result is that the emitted luminosities shown in Fig. \ref{fig:mfluxall7} are all super-Eddington, although the re-radiation effect has been taken into account. So the sub-Eddington puzzle is not solved for these simulations.

\begin{table*}
\caption{Properties of Outflow}
\label{tab:outflow}
\begin{tabular}{lcccccc}
\hline
Model & $\dot M_{out}(r_o)$& $v_r(r_o)$  &  $\dot p_w(r_o)$     &   $\dot E_{k}(r_o)$ & $\dot E_{th}(r_o)$      \tabularnewline
      & $(10^{25}~{\rm g~s^{-1}})$    & $({\rm km~s^{-1}})$     & $(10^{33}~{\rm g~cm~s^{-2}})$ &  $(10^{40}~{\rm ergs~s^{-1}})$ & $(10^{40}~{\rm ergs~s^{-1}})$ \tabularnewline
\hline
R5c  &  24.28(10.67)    &  800    & 21.57(5.72)      & 174.16(48.80)     & 3.77(0.83)      \tabularnewline
R6c  &  39.10(12.13)    &  1000   & 43.03(16.81)     &  547.17(459.39)   & 7.56(4.09)      \tabularnewline
R7c  &  127.04(11.80)   &  700    & 98.00(5.52)      & 1098.91(37.47)    & 12.31(0.63)      \tabularnewline
R8c  &  28.94(44.12)    &  600    & 23.58(64.40)     & 1110.45(10393.40) & 91.55(1206.77)     \tabularnewline
\hline
\end{tabular}
\\
\noindent \leftline{Note:~The values in brackets are the normalized standard deviations $\sigma_n$ of the time series values. For the Column $v_r(r_o)$, $v_r(r_o)$ is}
\leftline{the time-averaged mass flux-weighted radial velocity of outflow at $r_o$ (see panel c of Fig. \ref{fig:outflow}).
To avoid the influence of the outer}
\leftline{boundary, we actually calculate all the quantities at $\sim 6$ pc.}
\end{table*}

\subsection{Dependence on gas density at the outer boundary}

We can see from Table \ref{tab:model-summary} that the significance of re-radiative effects depends on the outer boundary density $\rho_0$. For example, when $\rho_0=5\times 10^{-20} {\rm g~cm^{-3}}$, the time-averaged luminosity decreases by a factor of 2 from model K7a to model R7a. The outflow rate at the outer boundary increases by a factor of about 7 from model K7a to model R7a. But when $\rho_0=10^{-20} {\rm g~cm^{-3}}$, the luminosity decreases by only $\sim 35\%$, and the outflow rate increases by a factor of about 2.5 from model K5a to R5a. Usually we find that the smaller the gas density at the outer boundary is, the weaker the re-radiation effect will be. This result is easy to understand. The thickness of the accretion disk is determined by the vertical component of the gravitational force, gradient of gas pressure, and the re-radiation force. The former two forces are proportional to density while the re-radiation force is proportional to the square of density (refer to equation \ref{eq:reradiation}). Therefore, the disk will become thicker and more exposed to the irradiation from the central AGN.

\subsection{Properties of outflows} \label{subsec:outflow}

\begin{figure*}
\begin{center}
\includegraphics[width=15.cm]{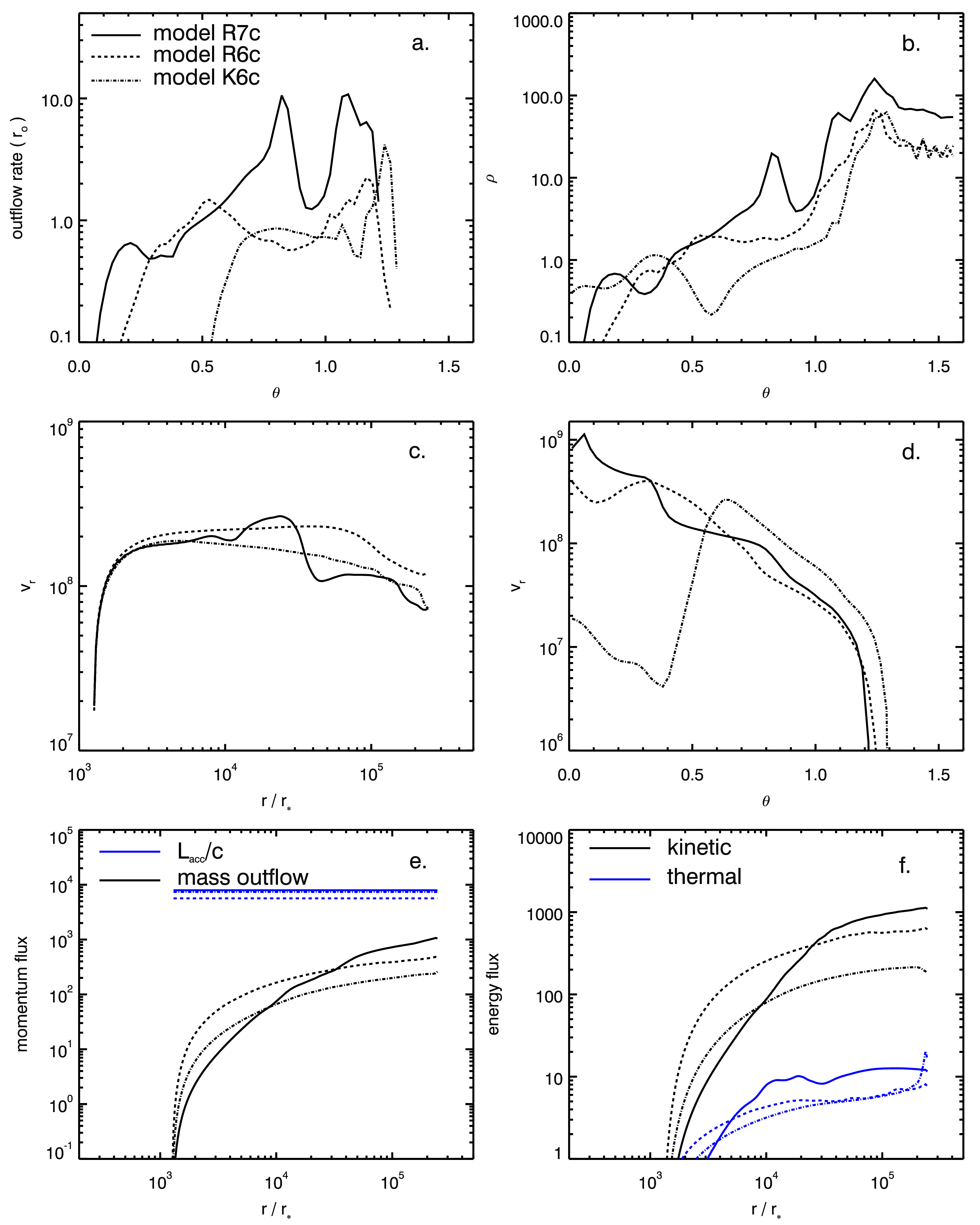}
\end{center}
\caption{Properties of outflow in models R7c (the solid lines), R6c (the dashed lines) and K6c (the dot-dashed lines). {\it Top}: The angular distribution of mass outflow rate (panel a, in unit of ${\rm 10^{25}~g~s^{-1}}$, defined as $\dot M_{out}(\theta)= 2\pi r^2\rho$~max$(v_r,0)sin(\theta)\Delta\theta$) and density (panel b, in unit of $10^{-21}~{\rm g~s^{-1}}$) at $\sim 6$ pc;
{\it Middle}: The radial profile of mass flux-weighted radial velocity of outflow (panel c, in unit of ${\rm cm~s^{-1}}$) and angular distribution of outflow velocity (panel d, in unit of ${\rm cm~s^{-1}}$) at $\sim 6$ pc;
{\it Bottom-left}: The radial profile of the momentum flux of outflow ($\dot p_w$, in unit of $10^{32}{\rm~g~cm~s^{-2}}$; the black lines) and radiation flux from the central AGN ($L_{\rm acc}/c$; the blue lines). {\it Bottom-right}: The radial profile of  the kinetic ($\dot E_k$; the black line) and thermal ($\dot E_{th}$; the blue lines) energy fluxes in unit of $10^{40}~{\rm ergs~s^{-1}}$. All the lines in this figure are time averaged.}
\label{fig:outflow}
\end{figure*}

In this section, taking models R5c, R6c, K6c, R7c and R8c as examples, we study the properties of the outflows. This is a sequence of increasing outer density and larger and lager rates of nominal inflow. These properties could be used to compare with observations. Some main properties are listed in Table \ref{tab:outflow}. From the left to right columns, they are the mass flux, radial velocity, momentum flux, kinetic energy flux, and thermal energy flux of outflow at the outer boundary, respectively.

Fig. \ref{fig:outflow} shows some additional properties of outflow for models R7c (the solid lines), R6c (the dashed lines) and K6c (the dot-dashed lines). Panels (a) and (b) show the angular distribution of mass fluxes and densities at $r_o$, respectively. For model R7c, we can see that most of the outflow occurs close to the disk surface, in the range of $35^{\circ}<\theta<68^{\circ}$. In the range of $\theta>68^{\circ}$ it is the inflowing region and there is almost no outflow. Panel (c) shows the radial profiles of the mass flux weighted radial velocities of outflows. The typical value of $v_r$ is $10^8{\rm cm~s^{-1}}$. We can see that in most of the radial range, the radial velocity does not keep increasing as we naively expect because of the acceleration by the radiation force. It even decreases somewhat at large radii. This is because of the continuous injection of new gas into the outflow. Panel (d) shows the angular distribution of $v_r$ at $r_o$. It monotonically increases from $\la 10^7{\rm cm~s^{-1}}$ close to the surface of inflowing region at $\theta\sim 68^{\circ}$ to $\sim 10^9{\rm cm~s^{-1}}$ close to the axis for models R7c and R6c. The dot-dashed line (model K6c) sharply decreases at high-latitude (small $\theta$) region. This is because some inflowing clumps often appear in that region which block the outflow (cf. the top panels of Fig. \ref{fig:contour7}, model K6c has similar structure). Due to the same reason, $\dot M_{out}(r_o)$ of model K6c is very weak at high-latitude region (see the dot-dashed lines in panel (a)). The black and blue lines in panel (e) show the radial profiles of the momentum fluxes of outflow ($\dot{p}_w$) and radiation from the central AGN, respectively. In panel (f), the solid and dotted lines respectively represent the radial profiles of the kinetic ($\dot{E}_k$) and thermal energy fluxes ($\dot{E}_{\rm th}$) of outflows.
Their definitions are:
\begin{equation}
   \dot p_w(r)=4\pi r^2\int^{\pi/2}_0 \rho v^2_r {\rm~sin}(\theta)~d\theta~~~{\rm for}~v_r > 0,
   \ \ \ \ \ \ \ \ \
\end{equation}
\begin{equation}
   \dot E_k(r)=2\pi r^2\int^{\pi/2}_0 \rho v^3_r {\rm~sin}(\theta)~d\theta~~~{\rm for}~v_r > 0,
   \ \ \ \ \ \ \ \ \
\end{equation}
\begin{equation}
   \dot E_{th}(r)=4\pi r^2\int^{\pi/2}_0 e v_r {\rm~sin}(\theta)~d\theta~~~{\rm for}~v_r > 0.
   \ \ \ \ \ \ \ \ \
\end{equation}
We see that the momentum flux $\dot{p}_w$ keeps increasing outwards, because more and more momentum of the radiation is transferred into the outflow gas. But we note that even at $r_o$, the momentum flux of outflow is still several times smaller than that of radiation. It will be interesting to adopt a larger $r_o$ to see how the momentum flux of outflow can finally become closer to that of the radiation. From Panel (f) we see that most of the power of outflow is in the form of kinetic energy rather than thermal energy. The highest power is reached at $r_o$, which is $\sim 1.1\times 10^{43} {\rm ergs~s^{-1}}$ for model R7c. As a comparison, for this model the luminosity from the central AGN is $1.8L_{\rm Edd}\sim 2.3\times 10^{46} {\rm ergs~s^{-1}}$. This is about three orders of magnitude higher than the kinetic power of outflow. Compared to the model of KP09, the efficiency of transferring the radiation into kinetic power of outflow is one order of magnitude higher. But it is still much lower than $\sim 0.05$, which is often assumed in cosmological simulations (see discussions in Kurosawa, Proga \& Nagamine 2009).

Comparing model R6c and K6c, we can see that model R6c has higher mass outflow rate at high-latitude region at $r_o$ and higher kinetic and thermal power at all radii. The outflow velocity  is also slightly higher. Compared with model R6c, mass outflow rate of model R7c is higher most obviously at the moderate-latitude region at $r_o$ as mentioned earlier. This is due to the increase of density at that region (see panel (b)). Compared to model R6c, the luminosity of model R7c is higher hence the radiative force larger, so  outflow stronger at small radii.

To better understand the overall energetics of the flow, we calculate the thermal energy of the input gas, the released gravitational energy, the energy absorbed from the center engine, the energy advected inside $r_i$, the energy advected outside $r_o$ and the energy re-emitted out per unit time for model R7c. We find that the first term is the smallest, $\sim10^{40}{\rm ergs~s^{-1}}$. The second term is comparable to the fourth term, $\sim3\times10^{43}{\rm ergs~s^{-1}}$. The fifth term ($\sim10^{43}{\rm ergs~s^{-1}}$) is much smaller than the third and sixth terms ($\sim6\times10^{44}{\rm ergs~s^{-1}}$). In other words, the energy absorbed from the center is comparable to the energy re-emitted out.

The value of $\eta_w$ (defined in equation \ref{eq:eta_w}) is given in the last column of Table \ref{tab:model-summary}. We can see that it becomes significantly larger when the re-radiation effect is included, $\eta_w\ga 1$. This implies that outflow rate is larger than the accretion rate. It is interesting to note that this is also the case for hot accretion flow (Yuan, Wu \& Bu 2012) and found by Li et al. (2013). Of course, in that case, the mechanism of producing outflow is completely different.

\begin{figure*}
\begin{center}
\includegraphics[width=15.cm]{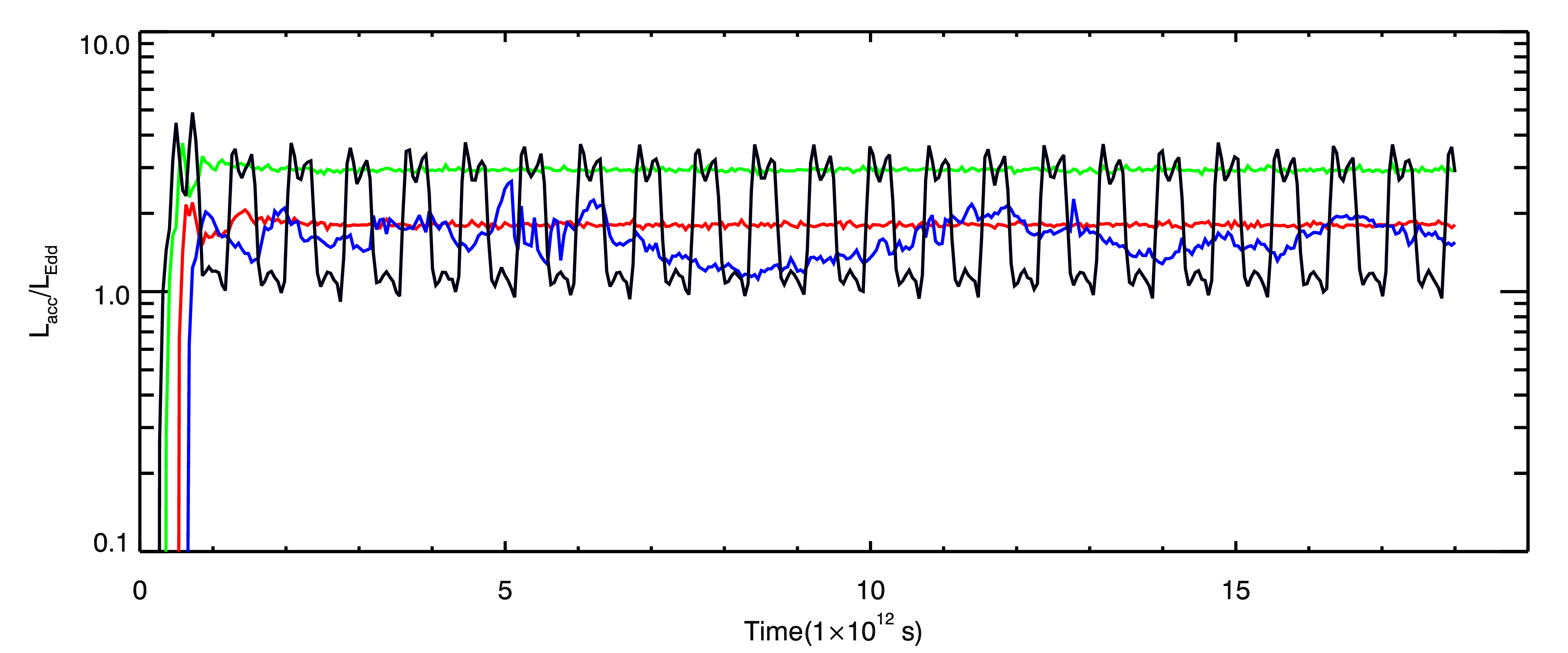}
\includegraphics[width=15.cm]{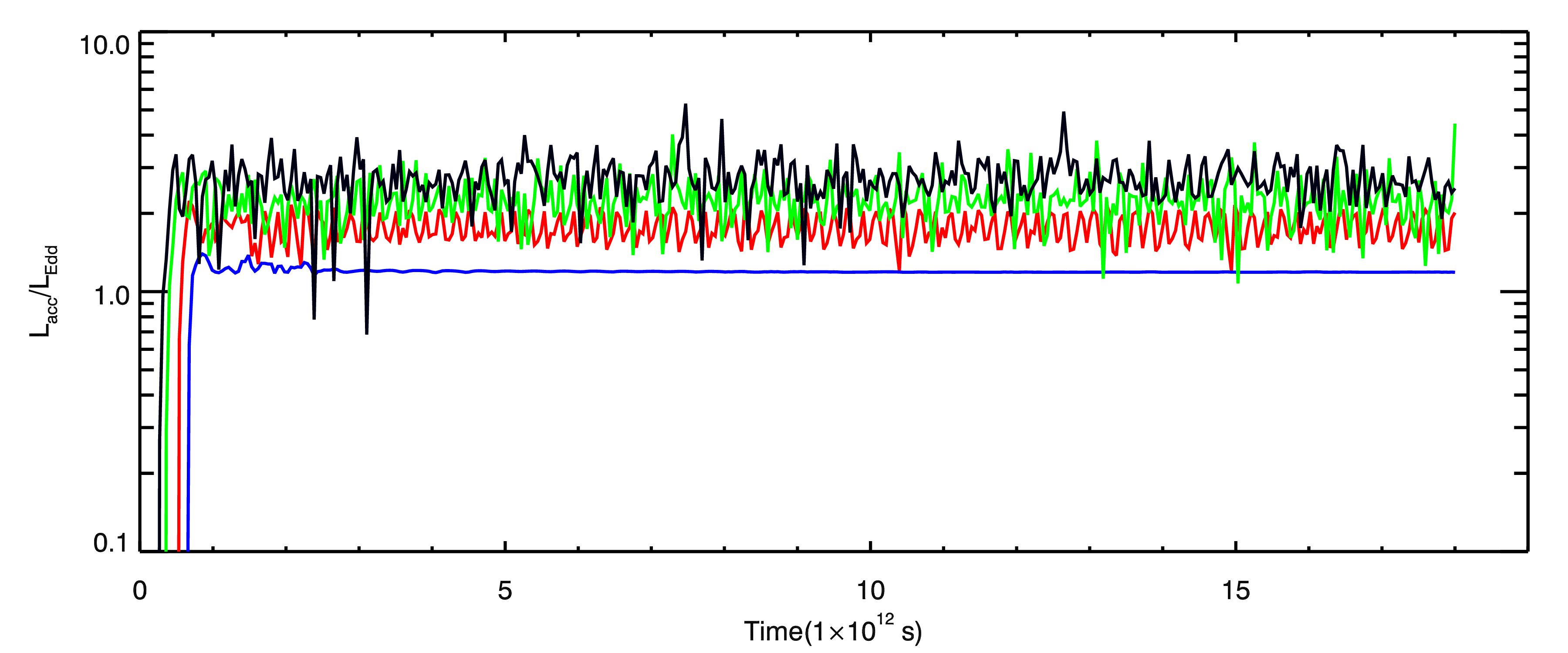}
\end{center}
\caption{Top panel: light curves for models with different $T_0$. The red, blue, green and black lines stand for models with $T_0=2\times10^6$ K, $5\times10^5$ K, $6\times10^6$ K and $10^7$ K, respectively. The four corresponding models are R7c (for comparison), R7c-lT, R7c-hT and R7c-hT2. Bottom panel: same as the top panel but for $\kappa_X=0$. The red, blue, green, and black lines represent models R7c-X0, R7c-lTX0, R7c-hTX0, and R7c-hT2X0, respectively. See Table \ref{tab:model-summary}.}
\label{fig:fig6}
\end{figure*}

\section{Discussion} \label{sec:discu}

\subsection{Effect of changing the temperature at the outer boundary}
So far we have fixed $T_0=2\times10^6$ K. It will be interesting to see the effect of varying $T_0$. With $T_0=2\times10^6$ K, the Bondi radius is $r_B\equiv GM_{BH}/c^2_s =4.8\times10^{19}{\rm~cm}\simeq 2.2~r_o$. If we adopt higher $T_0$ but maintain the outer boundary unchanged, the Bondi radius will be within our computational domain. This is an advantage since Bondi radius is the place where the inflow rate is determined (Li, Ostriker \& Sunyaev 2013). We choose three $T_0$ values: $5\times10^5$ K, $6\times10^6$ K and $10^7$ K. The corresponding $r_B$ of the two latter cases is smaller than $r_o$ now. These models with new $T_0$ are listed in the end of Table \ref{tab:model-summary} (Models 26-31).

The top panel of Fig. \ref{fig:fig6} gives the {\it emitted} luminosity $L_{\rm acc}$ as a function of time for Models R7c (red line; $T_0=2\times 10^6$ K), R7c-lT (blue line; $T_0=5\times10^5$ K); R7c-hT (green line; $T_0=6\times10^6$ K), and R7c-hT2 (black line; $T_0=10^7$ K). It is interesting to note that for model R7c-hT2 which has the highest $T_0$ and smallest $r_B$ ($\simeq0.43r_o$), the light curve oscillates regularly. Similar behavior has been found before such as the feedback study in elliptical galaxies by Novak, Ostriker \& Ciotti (2011). Such an oscillation is caused by radiative feedback. For model R7c-hT2, at some time the strong radiative heating heats the gas at a certain radius ($r_{\rm virial}$) above the local virial temperature. Then the accretion is partly suppressed since the gas becomes unbound. The accretion rate will thus begin to decrease, and subsequently the radiation from the central AGN will decrease. This makes the radiative heating weaker and the temperature of the gas will decrease until below the virial value. So the gas supply then recovers and the accretion becomes active again. The timescale of such a cycle should be determined by the accretion timescale at $r_{\rm virial}$. From the top panel of Fig. \ref{fig:fig6}, we see that the timescale is $\sim 10^{12}{\rm s}$. This is roughly the free-fall timescale at the outer boundary $r_o$, which is also the accretion timescale for our low angular momentum accretion.

We would like to emphasize that for such an oscillation behavior to occur, it is not necessary to have $r_{\rm virial}\ga r_B$. In fact, Yuan, Xie \& Ostriker (2009) find the same oscillation result even when $r_{\rm virial}< r_B$ when they consider the global Compton heating effect in hot accretion flows. The reason why the other three models do not have such oscillation is perhaps as follows. One is that the luminosity of the active phase from the other three models is smaller (refer to Fig. 6), thus the radiative heating is weaker. Another reason is that the specific thermal energy of the gas in Model R7c-hT2 is the highest among the four models at the same radius, therefore it is the easiest to be heated to be above the virial temperature.

In this context we would like to mention that the peak luminosity found in Yuan, Xie \& Ostriker (2009) is only $\sim 2\% L_{\rm Edd}$, much smaller than that in Model R7c-hT2, which is super-Eddington. One of the main reasons for the discrepancy is that the emitted spectrum from the innermost accretion flow is different. In the current work we assume a typical quasar spectrum, i.e., $T_X\sim 8\times 10^7K$ and $f_*=0.05$. Such a spectrum is likely produced by a standard thin disk plus a corona. However, in Yuan, Xie \& Ostriker (2009), the accretion flow is hot (such as advection-dominated accretion flow) thus the emitted spectrum is quite different. The radiation temperature of the spectrum in that case is $T_X\sim 3\times 10^9K$ and $f_*\sim 0.6$. In this case the Compton heating will become much stronger (e.g., Proga 2007a). In addition, the angular distribution of radiation is also different. In the case of hot accretion flow, the radiation is nearly spherically symmetric. This again increases the heating close to the equatorial plane. We have run some test simulations and confirmed the above effects. And lastly, the effectiveness of Compton heating is proportional to density of the flow. The specific angular momentum of the accretion flow in Yuan et al. (2009) is much larger than that in the current work. Therefore, for the same density, the corresponding accretion rate and therefore the luminosity of Yuan et al. (2009) will be significantly smaller than in the current work.

\subsection{Effect of X-ray optical depth}

So far we have treated the Thompson scattering opacity for the calculation of X-ray optical depth as if it were absorption. However, as we have mentioned above, this is likely an over-estimation (we neglect X-ray absorption). In this section, we want to check the optical depth effect of the outward propagation of X-ray from the central source by assuming $\kappa_X=0$. But in the calculation of the force we still keep the first term in equation (\ref{eq:scv}) for comparison purpose. The last few models labeled with `X0' in Table \ref{tab:model-summary} are such models.
\begin{figure*}
\begin{center}
\includegraphics[width=15.cm]{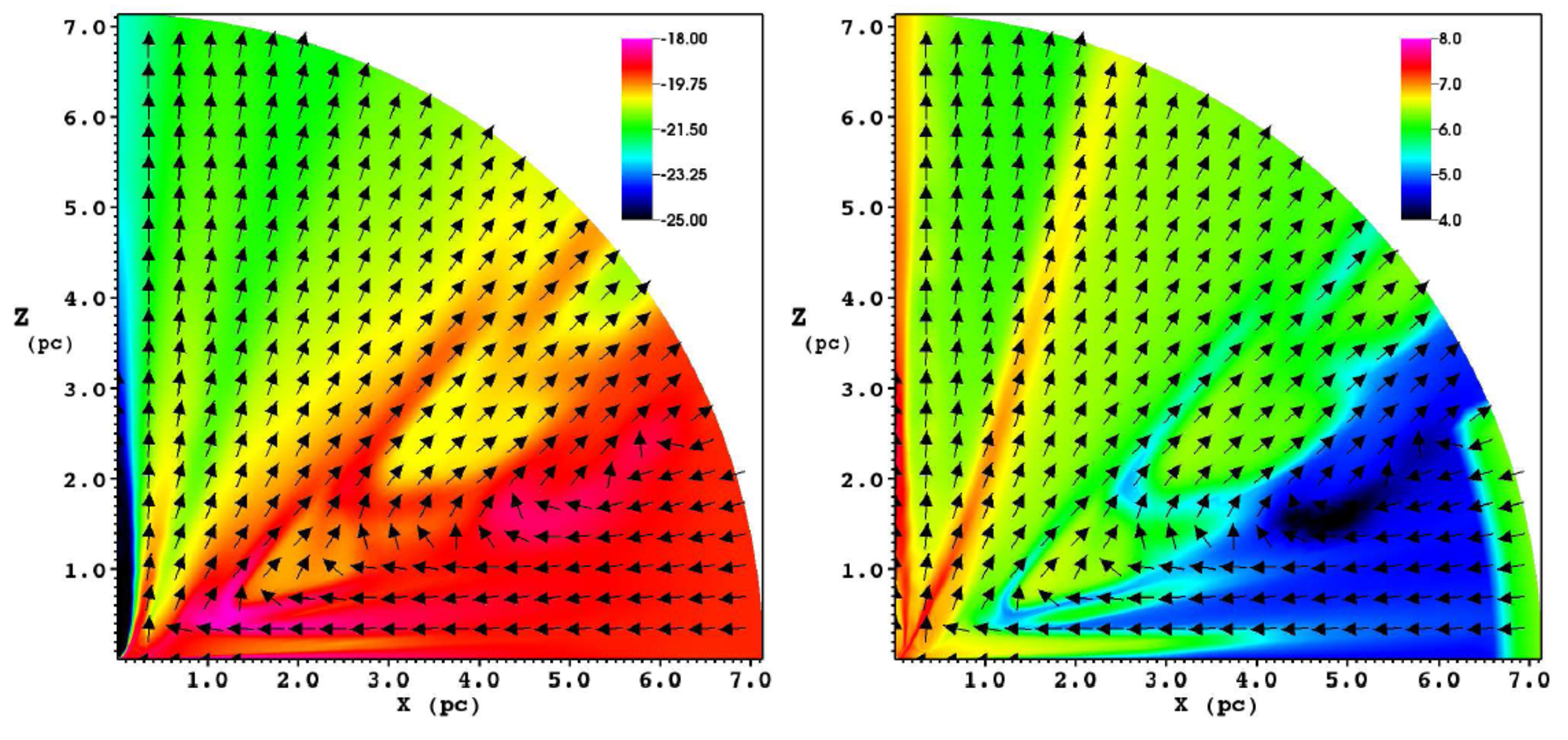}
\includegraphics[width=15.cm]{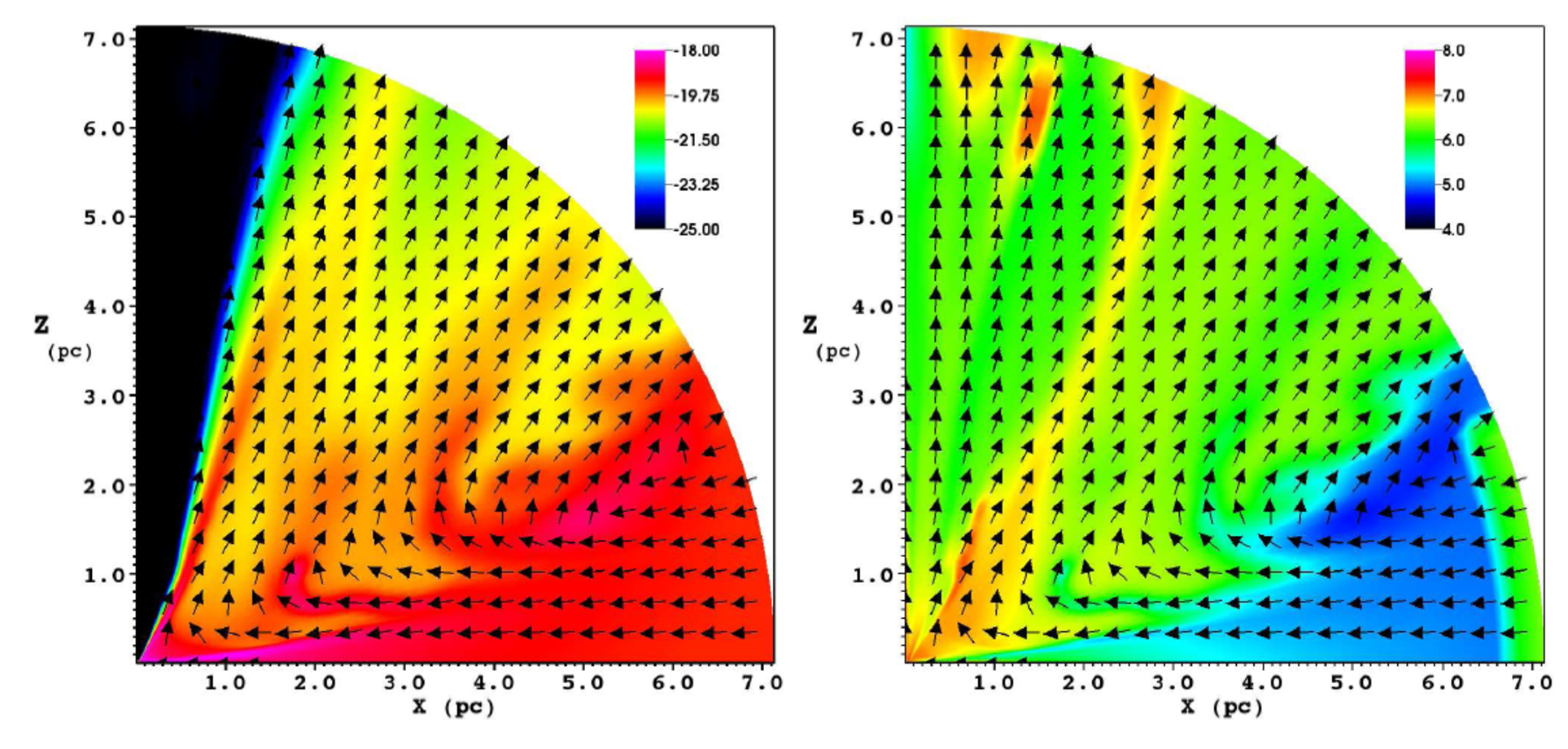}
\end{center}
\caption{Comparison of the logarithmic density (left) and temperature (right) contours between Model R7c (top panels) and Model R7c-X0 (bottom panels; with $\kappa_X=0$). The arrows represent the direction of velocity vector but not its magnitude.}
\label{fig:fig7}
\end{figure*}

Fig. \ref{fig:fig7} compares the logarithmic density and temperature contours with (model R7c, top panels) and without (model R7c-X0, bottom panels) X-ray opacity. When $\kappa_X=0$, there will be no X-ray attenuation. In this case, the ionization parameter $\xi$ increases and X-ray heating is enhanced especially at large radii (cf. equations \ref{eq:Gx}-\ref{eq:Gcomp}). We can see from the bottom-right panel of the figure that the gas temperature at low latitudes increases significantly. The gas density near the pole decreases greatly as the X-ray pushes the gas away. Once the gas density becomes lower, the X-ray heating becomes less effective, hence the temperature decreases. Note in table \ref{tab:model-summary} the high values of $\eta_w$ (cf. equation \ref{eq:eta_w}) in the models having $\tau_X=0$ and $r_o>r_B$. The outflow rates exceed the net accretion rates.

The top panel of Fig. \ref{fig:fig6} shows the light curves of various models which have been shown in the top panel of the same figure, except here $\kappa_X=0$. In the top panel, only the model with the highest $T_0$ (model R7c-hT2) oscillates. But here with the exception of model R7c-lTX0 which has the lowest $T_0$, the light curves of all other three models oscillate. The discrepancy is because of the enhanced X-ray heating. Comparing model R7c-hT2 in the top panel and model R7c-hT2X0 in the bottom panel, we can see that the timescale of the variability becomes smaller. This indicates that the virial radius $r_{\rm virial}$ becomes smaller. Another effect is that the amplitude of oscillation of model R7c-hT2X0 becomes smaller now. This is perhaps because a smaller fraction of the gas located at $r_{\rm virial}$ is heated to be unbound now.




\subsection{Observational implications}
We now briefly discuss the observational implications of our results. The first issue is the  origin of outflow observed in AGNs. They have been widely observed in the form of absorption lines in a variety of AGNs, including radio-quiet and radio loud, Seyfert and quasars, low-luminosity or high-luminosity ones (see Crenshaw, Kraemer \& George 2003 for a review). Especially, recently Tombesi et al. (2010, 2011, 2012) detected ultrafast outflows with velocities up to $(0.03-0.4)c$ via Fe K-shell absorption lines. They found that the ultrafast outflows are located in the interval $\sim(0.0003-0.03){\rm~pc}~\sim (10^2-10^4)~r_s$ from the black hole on average. The mass outflow rates are constrained between $\sim (0.01-0.1){\rm~M_\odot~yr^{-1}}$ and the average lower and upper limits on the mechanical power are $\dot E_k\simeq 10^{42.6}$ and $10^{44.6}{\rm~ergs~s^{-1}}$, respectively. These features are difficulty to explain with our simulations. In our models, the highest outflow velocity is about $0.03c$ (see panel (d) of Fig. \ref{fig:outflow}), and the outflows starts from $\sim 10^5~r_s$ (see Fig. \ref{fig:mfluxall7}), the mass outflow rates and outflow kinetic power at the outer boundary are $\gtrsim 4{\rm~M_\odot~yr^{-1}}$ and $\dot E_k\simeq (10^{42}-10^{43}){\rm~ergs~s^{-1}}$ (see Table \ref{tab:outflow}). However, based on these results we can not rule out the ``line-driven'' models as the origin of ultrafast outflow. This is because our simulations concentrate only on the large scale of the accretion flow, $(0.01\sim 7)$ pc, the innermost accretion flow is not included and substantial radiative acceleration may occur within $r_i$ (e.g., PSK00). We did adopt different inner boundary $r_i$ and found that the properties of outflow, including the location of origin, remain largely unchanged. This is likely because the angular momentum of the accretion flow we consider is rather small so there is no real disk formed in our simulation domain. On the other hand, the discrepancies between our simulation results and observations indicate that at least the ultrafast outflow must come from the innermost region of the accretion disk.

\section{Summary} \label{sec:summary}

In this paper we study the dynamics of accretion flow within $\sim 0.01 - 7$ pc from the central super-massive black hole irradiated by the central AGN. We focus on the outflow driven by the radiation via the Compton scattering and the line force. Compared to the previous works, the main improvement is that we include the re-radiation force, which is the additional force produced by the scattered photons and the reprocessed photons. We find that the results change significantly. The accretion flow now becomes thicker and is thus more exposed to the radiation from the central source. Consequently, the outflow rate measured at the outer boundary increases by about one order of magnitude. We find that the correlation between the Eddington ratio $\Gamma$ and the outflow rate $\dot{M}_{out}$ originally found in previous works still exists. The properties of  the outflows are calculated and compared to observations, such as the ratio of the outflow rate and the accretion rate. We find that some observed outflows can not be explained by this model, but note that the inferred high ratios of $\dot M_{out}/\dot M_{acc}$ occur naturally in our high Eddington ratio models. When the density of the gas at the outer boundary is sufficiently high, we still find super-Eddington accretion, even though we include the re-radiation effect and the outflow becomes stronger. We also investigate the effect of different temperatures at the outer boundary ($T_0$). We find that when $T_0$ is large ($\sim 10^7$ K), the emitted luminosity from the accretion flow oscillates significantly because of the radiative heating feedback.

One of the main aims of the present research is the sub-Eddington puzzle. Observations show that most AGNs are sub-Eddington, while on the other hand both analytical theory and numerical simulations of accretion flows have clearly shown that super-Eddington accretion can exist. In this paper, we hope to investigate whether the feedback at larger scale can solve the puzzle, especially after the re-radiation effect is taken into account. Unfortunately, we find that the super-Eddington accretion still exists (refer to Fig. \ref{fig:fig6}) but at a reduced rate.

There are some caveats in our study. These issues can be improved in the future, which may be helpful to solve the sub-Eddington puzzle. The first one is that we need to consider the probable large angular momentum of the accretion flow. When the angular momentum is large, we must include viscosity and a disk will be formed. In this case, such a disk itself will emit strong radiation, which may help to drive an outflow. Moreover, when the angular momentum is large, a strong outflow will be produced intrinsically by other mechanisms (Yuan, Bu \& Wu 2012; Bu et al. 2013). For example, outflow can be produced by buoyancy or magnetocentrifugal force in a hydrodynamical or magnetodynamical accretion flow, respectively (Yuan, Bu \& Wu 2012; see also Bai \& Stone 2013). The second caveat is that in the current simulations,  we only consider the radiation from within $r_i$ and we do not investigate the dynamics within $r_i$. However, strong outflows have been shown to exist from the innermost accretion flow and they will interchange their momentum and energy with the gas. So the feedback of the outflow should also be included (e.g., Ostriker et al. 2010; Novak et al. 2011; Choi et al. 2012). All the above-mentioned factors are expected to be able to make the outflow stronger and thus the inflow rate at $r_i$ and luminosity smaller. In addition, a simplified approach is adopted when we deal with the interaction between the radiation, from both the central force and re-radiation, and the gas.  Obviously a proper radiation transfer calculation is desired. And lastly, the opacity due to dust can also potentially greatly enhance the outflow (e.g., Novak et al. 2012; Roth et al. 2012; Faucher-Gigu\`{e}re, Quataert \& Hopkins 2013).

The values of $\eta_w>1$ found in our high inflow models are consistent with recent assessments of Moe et al. (2009), Dunn et al. (2010) and the references therein that winds carry out more mass in bright quasars than is being accreted on the central black holes.

\section*{Acknowledgments}

FY thanks the hospitality of Princeton University where part of the work was done. We thank Daniel Proga and Ryuichi Kurosawa for useful discussions/comments, and the anonymous referee for constructive comments. This work was supported in part by the Natural Science Foundation of China (grants 11103059, 11121062, and 11133005), the National Basic Research Program of China (973 Program 2009CB824800), and the CAS/SAFEA International Partnership Program for Creative Research Teams. The simulations were carried out both at Shanghai Supercomputer Center and the Super Computing Platform of Shanghai Astronomical Observatory.


\label{lastpage}

\end{document}